\documentclass[twocolumn,pra,amsmath,amssymb,floatfix,superscriptaddress]{revtex4-1}

\usepackage{graphicx}
\usepackage{epstopdf}
\usepackage{dcolumn}
\usepackage{bm}
\usepackage{amssymb}
\usepackage{setspace}
\usepackage{hyperref}
\usepackage{color}
\hypersetup{%
  pdfpagemode=None, 
  pdfstartpage=1,
  pdfstartview=FitH,
  pdfmenubar=true,
  pdftoolbar=true,
  colorlinks = true,
  linkcolor=blue,
  citecolor=blue,
  bookmarksopen=false
}

\begin{document}

\title{Controlling ultracold $p$-wave collisions with non-resonant light: Predictions of an asymptotic model for the generalized scattering volume}

\author{Anne Crubellier}
\email{anne.crubellier@u-psud.fr}
\affiliation{Laboratoire Aim\'e Cotton, CNRS, Universit\'e Paris-Sud,
  Universit\'e Paris-Saclay, ENS Cachan, Facult\'e des Science
  B\^atiment 505, 91405  Orsay Cedex, France} 
\email{anne.crubellier@u-psud.fr}

\author{Rosario Gonz\'alez-F\'erez}
\email{rogonzal@ugr.es}
\affiliation{Instituto Carlos I de F\'{\i}sica
  Te\'orica y Computacional and Departamento de F\'{\i}sica At\'omica,
  Molecular y Nuclear, Universidad de Granada, 18071 Granada,
  Spain}

\author{Christiane P. Koch}
\email{christiane.koch@uni-kassel.de}
\affiliation{Theoretische Physik,  Universit\"at Kassel,
  Heinrich-Plett-Str. 40, 34132 Kassel, Germany}

\author{Eliane Luc-Koenig}
\email{eliane.luc@u-psud.fr}
\affiliation{Laboratoire Aim\'e Cotton, CNRS, Universit\'e Paris-Sud,
  Universit\'e Paris-Saclay, ENS Cachan, Facult\'e des Science
  B\^atiment 505, 91405  Orsay Cedex, France}

\date{\today}
%::::::::::::::::::::::::::::::::::::::::::::::::::::::::::::::::::::::::::::::
\begin{abstract}
Interactions in a spin-polarized ultracold Fermi gas are governed by
$p$-wave collisions and can be characterized by the $p$-wave
scattering volume. Control of these collisions by Feshbach resonances
is hampered by huge inelastic losses. Here, we suggest non-resonant
light control of $p$-wave collisions, exploiting the anisotropic
coupling of non-resonant light to the polarizability of the
atoms. The $p$-wave scattering volume can be controlled by strong
non-resonant light, in close analogy to the $s$-wave scattering
length. For collision partners that are tightly trapped, the
non-resonant light induces an energy shift directly related to the
generalized scattering volume. This effect could be used to climb the
ladder of the trap. We also show that controlling the generalized
scattering volume implies control, at least roughly, over the
orientation of the  interparticle axis relative to the polarization
direction of the light at short interatomic distances. Our proposal is
based on an asymptotic model that explicitly accounts for the anisotropic dipole-dipole interaction which governs the ultracold collision dynamics at long-range.
\end{abstract}
%::::::::::::::::::::::::::::::::::::::::::::::::::::::::::::::::::::::::::::::
\maketitle
\renewcommand{\arraystretch}{1.3}
%-------------------------------------------------------------------------------
\section{Introduction  }
\label{sec:intro}
%-------------------------------------------------------------------------------
%

Collisions of neutral atoms or molecules at very low temperatures are universally described by a single parameter --- the $s$-wave scattering length for bosons and unpolarized fermions or the $p$-wave scattering volume for spin-polarized fermions~\cite{Friedrich16}. This parameter is the central quantity of the pseudopotential technique, where the interaction between two particles is accounted for in an effective way through the introduction of contact potentials for each partial $\ell$-wave~\cite{Derevianko05,Idziaszek06}. The effective interaction in an $s$-wave (resp. $p$-wave) collision vanishes when the scattering length (resp. volume) goes to zero, and likewise it becomes infinite when the scattering parameter becomes infinite. The latter case  corresponds to the appearance of a bound state at threshold. The sign of the scattering parameter renders the interaction to be effectively attractive or repulsive, deciding for example about the stability of a Bose-Einstein condensate or a degenerate Fermi gas against collapse at large densities. 
%% add citation which discusses or demonstrates the stability issue
Given this prominence, it is not surprising that controlling the scattering length or scattering volume has long been a primary goal in quantum gas experiments.

Initial proposals to control ultracold collisions of neutral atoms
focused on near-resonant optical manipulation of the scattering
length~\cite{FedichevPRL96,SlavaJCP01}. This type of control is
universal since it only requires a suitable optical
transition. However, due to inevitable spontaneous emission losses in
near-resonant coupling schemes, magnetic field control of
Fano-Feshbach resonances has become the most widely employed method of
choice to control collisions, in particular for alkali
atoms~\cite{ChinRMP10}. It requires presence of a hyperfine manifold
and sufficiently broad resonances. However, for $p$-wave collisions,
enormous inelastic losses were observed near Fano-Feshbach
resonances that can only be suppressed in specific geometries~\cite{ZhangPRA04,SchunckPRA05,GuenterPRL05,GaeblerPRL07,InadaPRL08,WaseemPRA17}. Even more severly, species other than the alkalis, such as alkaline earth atoms or mixtures of alkali and alkaline earth atoms, either do not possess Fano-Feshbach resonances at all or their resonances are too narrow to be exploited in magnetic field control. These species are promising candidates for important applications such as optical clocks or quantum simulation. Near-resonant optical control schemes have therefore been revisited~\cite{BlattPRL11,YamazakiPRA13,KillianPRL13}, albeit with mixed success due to spontaneous emission losses. 

Spontaneous emission is minimized for non-resonant light control~\cite{GonzalezPRA12,TomzaPRL14}. Non-resonant light universally couples to the polarizability of the atoms, independent of the frequency of the light and the energy level structure of the atoms, as long as the frequency remains far detuned from any resonance. This interaction can be used to modify both shape and Fano-Fesbach resonances~\cite{GonzalezPRA12,TomzaPRL14,CrubellierNJP15a}. Moreover, for sufficiently high intensity, the non-resonant light coupling results in a variation of the scattering length with the field intensity~\cite{TomzaPRL14}, similarly to the control of the scattering length by a magnetic field near a Fano-Feshbach resonance~\cite{ChinRMP10}. 
This gives rise to non-resonant light control of the scattering length~\cite{CrubellierPRA17}. In particular, 
the scattering length diverges when, with increasing intensity, a shape resonance crosses the threshold to become bound or when the field-dressed potential becomes sufficiently deepened to accommodate an additional bound level~\cite{GonzalezPRA12,CrubellierNJP15a}. 
It is natural to ask whether this type of control can be extended to $p$-wave collisions of spin-polarized fermions. 

To answer this question, we employ an asymptotic model which replaces the interaction potential by its asymptotic part~\cite{GaoPRA98,GaoPRA01,GaoJPB03,CrubellierJPB06,GaoPRA09,LondonoPRA10}. This approximation is well justified at ultralow temperatures.  
When controlling a pair of atoms with non-resonant light, the
resulting asymptotic
Hamiltonian~\cite{CrubellierNJP15a,CrubellierPRA17} turns out to be
identical to the one describing the control of atom-atom interaction
by a static electric field~\cite{MarinescuPRL98} as well as that
describing ultracold collisions of polar
molecules~\cite{RoudnevJPB09,BohnNJP09}. These problems have in common
that they are all governed by the anisotropic dipole-dipole
interaction, which decreases with the interatomic separation as
$1/R^{3}$ and introduces a coupling between all partial $\ell$-waves
of the same parity. The crucial parameter of the corresponding
asymptotic model is the $p$-wave scattering volume which may, on first
glance, appear to be ill-defined in the presence of dipole-dipole
interaction. However, we have shown in the preceding paper, referred
to as Paper I~\cite{paperI}, how to remedy this problem by suitably generalizing the definition of the scattering volume. We can thus proceed now to examine non-resonant light control of the scattering volume that involves exactly this type of interaction. 

The present paper is organized as follows. 
Section~\ref{sec:dip-dip} recalls the asymptotic model for an interparticle interaction of dipole-dipole type in Sec.~\ref{subsec:hamil} and lists a few typical physical examples of this model in  Sec.~\ref{subsec:examples}. We use the asymptotic model to make general predictions for non-resonant light control of the scattering volume in Sec.~\ref{sec:predic-v}, distinguishing between weak and strong confinement in Secs.~\ref{subsec:weak} and \ref{subsec:traps}.
Section~\ref{sec:orient}  analyzes the connection between controlling the scattering volume and the orientation of the interparticle axis relative to the polarization direction for the pure $p$-wave case in Sec.~\ref{subsec:orient} and for multiple channels in Sec.~\ref{subsec:orient-multi}. 
We conclude in Sec.~\ref{sec:conclusion}. 

%
%-------------------------------------------------------------------------------
\section{Model}
\label{sec:dip-dip}
%
% %-------------------------------------------------------------------------------
% %
% %+++++++++++++++++++++++++++++++++++++++++++++++++++++++++++++++++++++++++++++++
\subsection{Hamiltonian and asymptotic Schr\"odinger equation}
\label{subsec:hamil}
%+++++++++++++++++++++++++++++++++++++++++++++++++++++++++++++++++++++++++++++++
%
The model describing the relative motion of two dipoles aligned along the laboratory $Z$-axis and interacting via a short range potential is close to the one described in our previous study~\cite{CrubellierPRA17}. For completeness, we briefly recall here the Hamiltonian and the reduced units that allow for a general treatment, independent of the specific parameters of the particles. In the Born-Oppenheimer approximation and employing spherical coordinates, the Hamiltonian reads
\begin{equation}
  \label{eq:2D_Hamil}
  H =   T_R+\frac{\hbar^2\mathbf{L}^2}{2\mu  R^2} +V_g(R)
    + {\cal D}  \frac{3\cos^2\theta -1}{R^3} \,,
\end{equation}
where $R$ denotes the interparticle separation and $\theta$ the angle between $\vec R$ and the  $Z$ axis.
$\mu$ is the reduced mass, $T_R$ the radial kinetic energy, $\mathbf{L}$ the orbital angular momentum operator, and $V_g(R)$ the potential describing the short-range interactions.
For simplicity, $V_g(R)$ is limited here to the van der Waals potential, $V_g(R)$=$-C_6/R^6$, with $C_6$ the van der Waals coefficient.
The last term in the Hamiltonian~\eqref{eq:2D_Hamil} stands for the
anisotropic dipole-dipole interaction governing the scattering
properties at large interparticle distance. This interaction can be
due to a non-resonant light with intensity $I$, linearly polarized
along $Z$ axis coupling to the polarizability anisotropy of the
particles. Equivalently, it can be caused by an electric or magnetic
field  along the $Z$ axis, coupling to corresponding aligned permanent dipole moments.
The equivalence is expressed in terms of the dipolar interaction strength $\mathcal D$,
\begin{equation}
  \label{eq:corresp}
 {\cal D}\,\leftrightarrow\,\frac{1}{4\pi\epsilon_0}d_1d_2\,\leftrightarrow\,\frac{\mu_0}{4\pi}m_1m_2\,\leftrightarrow\,\frac{4 \pi I}{c}\alpha_1\alpha_2\,,
\end{equation}
where $d_{1,2}$ ($m_{1,2}$) denotes the magnitude of the electric (magnetic) dipole moments, whereas $\alpha_{1,2}$ are the static polarizabilities of the two particles, with a dimension of volume~\cite{CrubellierNJP15a}. Here, $c$ denotes the velocity of light, $\epsilon_0$ the  permittivity of vacuum and $\mu_0$ the vacuum permeability.

The  Hamiltonian~\eqref{eq:2D_Hamil} commutes with parity and with $L_Z$, the projection of the orbital angular momentum on the laboratory $Z$ axis. As a result, the projection quantum number $m$ is conserved. Non-resonant light control of the scattering length concerning  $m$=0 and even-parity $\ell$ states has been discussed in Ref.~\cite{CrubellierPRA17}. Here, we consider odd-parity wave functions with $m$=0 or $\pm 1$.

A universal form of the Hamiltonian~\eqref{eq:2D_Hamil} is obtained by introducing reduced units. These can be chosen to eliminate the scaling factor of the rotational kinetic energy together with the prefactor of either the dipole-dipole interaction or the van der Waals term. In the latter case, hereafter referred to as 'van der Waals reduced units' (and denoted by ru), the reduced units of length $x$, energy ${\mathcal E}$, and non-resonant field intensity ${\mathcal I}$ are, respectively, defined by $R$=$\sigma x$, $E - E_0 $=$\epsilon \,{\mathcal E}$, where $E_0$ denotes the shift of the dissociation limit induced by the non-resonant light, and $I$=$\beta~{\mathcal I}$~\cite{LondonoPRA10,CrubellierNJP15a}. The corresponding characteristic length $\sigma$, energy $\epsilon$ and field intensity $\beta$ are equal to
\begin{subequations}
\label{eq:scaling}
\begin{eqnarray}
\sigma & = & \left(\frac{2\mu C_6}{\hbar^2}\right)^{1/4}\,, 
\label{eq:sigma}\\ 
\epsilon & = & \frac{\hbar^2}{2\mu\sigma^2}\,,
\label{eq:epsilon} \\ 
\beta & = & \frac{c}{12\pi} \frac{\hbar^{3/2}C_6^{1/4}}{\alpha_1\alpha_2(2\mu)^{3/4}} 
=\frac{c\sigma^3\epsilon}{12\pi\alpha_1\alpha_2}\,.
\label{eq:beta}
\end{eqnarray}
\end{subequations}
These unit conversion factors contain all the information specific to
the particle species, i.e., reduced mass $\mu$, van der Waals
coefficient $C_6$, and polarizabilities $\alpha_{1}$ and
$\alpha_{2}$. With these units, the asymptotic Schr\"odinger equation
for the wave function $f(x,\theta,\phi)$, where $\phi$ denotes the
azimuthal angle, becomes 
\begin{equation}
  \label{eq:asy}
  \left[-\frac{d^2}{dx^2} - \frac{1}{x^6} + \frac{\mathbf{L}^2}{x^2}
    - {\mathcal I}  \frac{\cos^2\theta -1/3}{x^3} - {\mathcal E}
  \right] f (x,\theta,\phi) = 0\,,
\end{equation}
where the van der Waals interaction is indeed described by the universal term $-1/x^6$. The non-resonant field intensity ${\mathcal I}$ is a tunable parameter allowing to control the collision. For a dipole-dipole interaction characterized by the strength $\mathcal D$, the reduced intensity is $\mathcal I$=$ 3\mathcal D /\epsilon \sigma^3$. 

The second set of reduced units, hereafter referred to as 'dipole-dipole units' (and denoted by ru(dd)), is obtained by introducing the characteristic length $D$ and energy $E_D$~\cite{BohnNJP09},  
\begin{subequations}
\label{eq:scalingdip}
\begin{eqnarray}
D & = & \frac{\mu}{\hbar^2}{\cal D}\,, 
\label{eq:D}\\
E_D & = & \frac{\hbar^2}{\mu  D^2} = \frac{\cal D}{D^3} \,,
\label{eq:ED} 
\end{eqnarray}
\end{subequations}
such that $R=D\overline{x}$ and $E=E_D\overline{\mathcal E}$.
In these reduced units, the asymptotic Schr\"odinger equation reads
\begin{equation}
  \label{eq:asydip}
  \left[-\frac{d^2}{d\overline{x}^2} - \frac{\overline c_6}{\overline{x}^6} + \frac{L^2}{\overline{x}^2}
    - 6  \frac{\cos^2\theta -1/3}{\overline{x}^3} - 2{\overline{\cal{E}}}
  \right] f (\overline{x},\theta,\varphi) = 0\,,
\end{equation}
where $\overline c_6$, the reduced strength of the van der Waals interaction, is given by 
\begin{equation}
  \label{eq:c6}
\overline c_6 =  2 \mu C_6/(\hbar^2 D^4) \,.
\end{equation}
Whereas in Eq.~\eqref{eq:asy}, the short-range van der Waals interaction is described by a universal term, it is the long-range dipole-dipole interaction which appears as universal in Eq.~\eqref{eq:asydip}. 
Converting the characteristic length and energy from one unit set to the other depends only on $\mathcal I$, 
\begin{subequations}
\label{eq:corresp-units}
\begin{eqnarray}
D & = & \frac{\mathcal I}{6}\,\sigma, 
\label{eq:D-sigma}\\ 
E_D & = & \frac{72 }{\mathcal I^2}\,\epsilon \, ,
\label{eq:ED-epsilon}
\end{eqnarray}
\end{subequations}
whereas the non-universal system-dependent parameters $\overline c_6$ and ${\mathcal I}$ in Eqs.~\eqref{eq:asydip} are related by
\begin{equation}
\label{eq:tunable}
\overline c_6=\frac{\sigma^4}{D^4}=\frac{6^4}{{\mathcal I}^4}\,.
\end{equation}
%
%
%+++++++++++++++++++++++++++++++++++++++++++++++++++++++++++++++++++++++++++++++
\subsection{Physical examples described by the asymptotic model}
\label{subsec:examples}
%+++++++++++++++++++++++++++++++++++++++++++++++++++++++++++++++++++++++++++++++
%
%
%%%%%%%%%%%%%%%%%%%%%%%%%%%%%%%%%%% TABLE I %%%%%%%%%%%%%%%%%%%%%%%%%%%%%%%%%%%%
% nrf
%%%%%%%%%%%%%%%%%%%%%%%%%%%%%%%%%%%%%%%%%%%%%%%%%%%%%%%%%%%%%%%%%%%%%%%%%%%%%%%%           
\begin{table*}[tb]
  \begin{tabular}{|c|c|c|c|c|c|c|c|}
    \hline
 pair& $C_6$  & $\alpha_1$, $\alpha_2$  & $\sigma$  & $\epsilon/k_B$  & $\beta$  & $m/\sqrt {\mathcal I}$  & $d/\sqrt{\mathcal I}$   \\
     &    ($a_0^6$) &    ($a_0^3$) &  ($a_0$) & ($\mu$K)  & (GW cm$^{-2})$ & ($\mu_B$) &(Debye) \\
\hline \hline
$^{88}$Sr$_2$       & 3248.97 & 186.25       & 151.053  & 86.365  & 0.6358 & 4.858 & 0.04506   \\ \hline
$^{86}$Sr-$^{88}$Sr & 3248.97 & 186.25       & 150.618  & 87.875  & 0.6413 & 4.879 & 0.04525   \\ \hline
$^{86}$Sr$_2$       & 3248.97 & 186.25       & 150.188  & 89.393  & 0.6468 & 4.900 & 0.04545   \\ \hline
$^{87}$Sr$_2$       & 3248.97 & 186.25       & 150.623  & 87.855  & 0.6412 & 4.879 & 0.04525   \\ \hline \hline
$^{171}$Yb$_2$      & 1932.   & 142.         & 156.639  & 41.303  & 0.5833 & 3.548 & 0.03290   \\ \hline
$^{172}$Yb$_2$      & 1932.   & 142.         & 158.868  & 40.943  & 0.5807 & 3.540 & 0.03283   \\ \hline
$^{173}$Yb$_2$      & 1932.   & 142.         & 157.096  & 40.588  & 0.5782 & 3.532 & 0.03276   \\ \hline
$^{174}$Yb$_2$      & 1932.   & 142.         & 157.323  & 40.238  & 0.5757 & 3.525 & 0.03269    \\ \hline \hline
$^{40}$K-$^{87}$Rb  & 4106.5  & 292.88 309.88& 142.284  & 156.282 & 0.3674 & 5.975 & 0.05541    \\ \hline
$^{7}$Li-$^{133}$Cs & 2933.8  & 163.98 402.20&  91.885  & 1539.41 & 1.3416 & 9.731 & 0.09025   \\ \hline
$^{87}$Rb-$^{133}$Cs& 5284.9. & 309.98 402.20& 178.379  & 51.802  & 0.1747 & 4.828 & 0.04478    \\ \hline \hline
$^{52}$Cr$_2$       & 733.    & 78.          & 91.2731  & 400.338 & 3.7071 & 4.913 & 0.04556   \\ \hline
$^{53}$Cr$_2$       & 733.    & 78.          & 91.7093  & 389.047 & 3.6545 & 4.878 & 0.04524    \\ \hline
  \end{tabular}
  \caption{\label{tab:nrf} Examples of atom pairs which are good candidates for collision control by non-resonant light together with their van der Waals constant $C_6$, atomic polarizabilities $\alpha_{1/2}$, taken from Ref.~\cite{Schwerdtfeger06}, and values for the reduced units of length $\sigma$, energy $\epsilon$ and intensity $\beta$, cf. Eq.~\eqref{eq:scaling}. The proportionality coefficients that relate the interaction with non-resonant light of intensity $\mathcal I$ to the dipole-dipole interaction of a pair with permanent magnetic $m$ or electric $d$ dipole moment, cf. Eq.~\eqref{eq:corresp}, are also given.}
\end{table*}
%
%%%%%%%%%%%%%%%%%%%%%%%%%%%%%%%%%%%%%%%%%%%%%%%%%%%%%%%%%%%%%%%%%%%%%%%%%%%%%%%%%
%
We summarize the values of the universal as well as system-dependent parameters for a few atoms and molecules to which our model applies, either when they interact with a non-resonant field, cf. Table~\ref{tab:nrf}, 
or when they interact with each other via a permanent electric or magnetic dipole moment, cf. Table~\ref{tab:dip-dip}. Table~\ref{tab:nrf} presents our selection of good candidates for control with non-resonant light out of the species that have already experimentally been cooled down to temperatures in the milli-kelvin or even nano-kelvin range. While all atomic or molecular collision partners are polarizable and thus interact with non-resonant light, the field strengths required for control are rather different. For the non-resonant light to significantly alter the scattering properties, the field-induced term in the Hamiltonian~\eqref{eq:2D_Hamil} needs to compete with the rotational kinetic energy. In other words, large polarizabilities and reduced masses are favorable, explaining our choice of strontium~\cite{MickelsonPRA10,StellmerPRA13} and ytterbium~\cite{FukuharaPRL07,KitagawaPRA08,CappelliniPRL14}. 
For even isotopes, these atoms have a closed shell ground state $^1S_0$ with vanishing total angular momentum $J$=0 and possess neither a permanent magnetic dipole moment nor a hyperfine manifold.
In addition to the atomic homonuclear pairs with no permanent electric
or magnetic dipole moment, we consider heteronuclear dialkali-metal
pairs with permanent electric dipole moment: the smallest
(KRb~\cite{NiSCIENCE08,OspelkausPRL08}), the largest (LiCs)~\cite{DeiglmayrPRA10} 
and an intermediate example (RbCs~\cite{ShimakasakiCPC16}).  Finally, we include the pair of transition metal atom Cr with atomic ground level $3d^5 4s$ $^7S_3$, with a large permanent magnetic dipole moment. 
For these pairs, the reduced length $\sigma$ characterizing the range of interatomic separation where the van der Waals interaction prevails is of the order of 100 to 200 a$_0$. The reduced energy $\epsilon$ is in the micro-kelvin range. The reduced unit of non-resonant light intensity, $\beta$, of the order of 1$\,$GW/cm$^2$, provides an estimate  for the intensity required to effectively control the collisions. While such a high intensity is challenging to realize experimentally, a tight focus is one way to reach it, as discussed in Refs.~\cite{TomzaPRL14,CrubellierPRA17} for the control of the $s$-wave scattering length.
Application of a non-resonant light of reduced intensity $\mathcal I$ is identical to dipole-dipole interaction in systems with a permanent electric $d$ or magnetic $m$ dipole moment, increasing as $\sqrt \mathcal I$ and proportional to $C_6^{1/8}/(\alpha \mu^{3/8}$), see Eq.~\eqref{eq:corresp}. For an intensity of $\mathcal I$=1, i.e., $I=\beta\,$GW/cm$^2$ (with $\beta$ evaluated from Eq.~\eqref{eq:beta}), the equivalent electric dipole moments reported in Table \ref{tab:nrf} are about 0.03 to 0.1 Debye,  whereas the equivalent magnetic dipole moments are in the range from 3.5 to 10 $\mu_B$. 
The pair RbCs (Cr$_2$) presents the largest (smallest) value for the product of the polarizabilities $\alpha_1 \alpha_2$ or, equivalently, the smallest (largest) reduced unit for the field intensity $\beta$. It is thus the most (least) 
favorable candidate for control by non-resonant light. Note that the very large values of the equivalent dipole moments for LiCs result from the very small reduced mass $\mu$.

%
%%%%%%%%%%%%%%%%%%%%%%%%%%%%%%%%%%% TABLE II %%%%%%%%%%%%%%%%%%%%%%%%%%%%%%%%%%%%%%%
% dipole-dipole
%%%%%%%%%%%%%%%%%%%%%%%%%%%%%%%%%%%%%%%%%%%%%%%%%%%%%%%%%%%%%%%%%%%%%%%%%%%%%%%%%%%%
%
\begin{table*}[tb]
  \begin{tabular}{|c|c|c|c|c|c|c|c|c|c|}
    \hline
 pair& $C_6$  & $m$ & $d$  & $\alpha$ & $D$  & $E_D/k_B$  & ${\overline c}_6$    & ${\mathcal I_c}$ & $\beta$  \\ 
& $(a_0^6)$ &  ($\mu_B$) & (Debye) & $(a_0)^3$ & ($a_0$) &($\mu$K) &(ru(dd)) & (ru)& (GW cm$^{-2})$ \\ \hline \hline
$(^{40}$K-$^{87}$Rb)$_2$   & 15972.  & -      & 0.566 & 602.86 & 5734.14  & $83.0502\cdot 10^{-3}$ & $3.41677\cdot 10^{-6}$  & 139.556 & 0.06862 \\ \hline
$(^{39}$K-$^{87}$Rb)$_2$   & 15972.  & -      & 0.566 & 602.86 & 5688.93  & $85.0461\cdot 10^{-3}$ & $3.49889\cdot 10^{-6}$  & 138.73  & 0.06903 \\ \hline
$(^{7}$Li-$^{133}$Cs)$_2$  & 4585400 & -      & 5.5   & 566.18 & 597139.  & $6.944\cdot 10^{-6}$  & $9.19856\cdot 10^{-12}$ & 3445.25 & 0.29758 \\ \hline
$(^{87}$Rb-$^{133}$Cs)$_2$ & 147260. & -      & 1.23  & 712.17 & 46917.3  & $716.021\cdot 10^{-6}$ & $1.21779\cdot 10^{-8}$  & 571.161 & 0.05674 \\ \hline \hline
$^{52}$Cr$_2$             & 733.    & 6.00696 & -     & 78. & 22.7413  & 12897.6           & 259.46             & 1.4944  & 3.7071 \\ \hline
$^{53}$Cr$_2$             & 733.    & 6.00696 & -     & 78. & 22.1792  & 12180.4           & 245.05             & 1.5165  & 3.6545  \\ \hline \hline
$^{161}$Dy$_2$           & 1890.   & 10.0046 & -     & 165. & 195.267  & 56.4625           & 0.381358           & 7.63516 & 0.44952 \\ \hline
$^{162}$Dy$_2$            & 1890.   & 10.0046 & -     & 165. & 196.663  & 55.3203           & 0.372951           & 7.67783 & 0.44744 \\ \hline
$^{164}$Dy$_2$            & 1890.   & 10.0046 & -     & 165. & 199.095  & 53.3178           & 0.35945            & 7.74893 & 0.44334 \\ \hline \hline
$^{167}$Er$_2$            & 1760.   & 7.00732 & -     & 153. & 99.3172  & 210.4070          & 5.5045             &  3.9172 & 0.49965 \\ \hline
$^{168}$Er$_2$            & 1760.   & 7.00732 & -     & 153. & 99.9123  & 206.6690          & 5.4067             &  3.9348 & 0.49742 \\ \hline\hline
$(^{168}$Er$_2)^2$        & 7040.   & 14.0046 & -     & 306. & 799.2990 & 1.61460           & 0.0105599          & 18.718  & 0.10457 \\ \hline 
  \end{tabular}
  \caption{\label{tab:dip-dip} Examples of pairs of molecules and atoms with notable dipole moment, either magnetic $m$ or electric $d$, and with polarizability $\alpha$ taken from Ref.~\cite{Schwerdtfeger06}. 
The reduced units of length $D$ and energy $E_D$ are specific to the dipole-dipole interaction, see Eq.~\eqref{eq:scalingdip}. 
The value of the van der Waals constant is $C_6$ given in atomic units
and ${\overline c}_6$ in dipole-dipole reduced units (ru(dd)), see Eq.~\eqref{eq:c6}.
A non-resonant field of intensity $\mathcal I_c$ in reduced units (ru)
specific to the van der Waals interaction, with conversion factor $\beta$, see Eq.~\eqref{eq:beta}, would mimic the effect of the permanent dipole moments, cf. Eq.~\eqref{eq:Ic}.
}
\end{table*}
%%%%%%%%%%%%%%%%%%%%%%%%%%%%%%%%%%%%%%%%%%%%%%%%%%%%%%%%%%%%%%%%%%%%%%%%%%%%%%%%%%%%
%
Table~\ref{tab:dip-dip} presents the reduced units of length $D$ and energy $E_D$,  cf. Eqs.~\eqref{eq:scalingdip}, for collision partners with a permanent electric or magnetic dipole moment, assumed to be aligned. It starts with pairs of  heteronuclear dialkali-metal molecules, namely pairs of KRb, LiCs and RbCs~\cite{NiNAT10,MirandaNATPHYS11}, in their lowest rovibrational level. These molecules possess a large permanent electric dipole moment varying from $d$=$0.56\,$D for KRb up to 5.5$\,$D for LiCs (see Table I of Ref.~\cite{LepersPRA13}). The polarizability of the diatomic molecule is taken to be equal to the sum of the polarizabilities of the two constituent atoms. For these pairs, the van der Waals interaction in the lowest rovibrational level is huge, three orders of magnitude larger than in a pair of alkali atoms (see Table II of Ref.~\cite{LepersPRA13}). However, the reduced strength of the van der Waals interaction $\overline c_6$ decreases as $\mu C_6 D^{-4}$, see Eq.~\eqref{eq:c6}.
Since the unit of length $D\propto \mu d^2$ is also very large, especially for LiCs ($D\sim 0.32 \mu$m due to large $d$) and for RbCs ($D\sim 0.25 \mu$m due to large $d$ and $\mu$),  $\overline c_6$ takes values between 10$^{-11}$ to $10^{-6}$. Therefore, the van der Waals interaction is almost negligible, and the dipole-dipole interaction governs the dynamics.

Table~\ref{tab:dip-dip} also presents homonuclear pairs of atoms, bosonic or fermionic, with a large total angular momentum $J$ and therefore a large permanent magnetic moment: pairs of the transition metal atom Cr, with atomic ground level $3d^5 4s$ $^7S_3$~\cite{PasquiouPRA10, NaylorPRA15}, pairs of the lanthanide atoms Dy~\cite{LuPRL12,TangPRA15} and Er~\cite{AikawaPRL14,FrischPRL15}, with respective atomic ground levels $4f^{10} 6s^2$ $^5I_8$ and $4f^{12} 6s^2$ $^6H_6$. In their lowest state $^{2S+1}L_J\, |M_J|$=$J$, with Land\'e factor $g_J$, these atoms possess a large permanent magnetic dipole moment $m$=$\mu_B g_J J$, and two collision partners strongly interact via magnetic dipole-dipole interaction. The van der Waals coefficients for Er and Dy are taken from Ref.~\cite{LepersPRA14} and Ref.~\cite{KotochigovaPCCP11}, respectively. 
Finally, Table~\ref{tab:dip-dip} considers the collision between two Er$_2$ molecules~\cite{FrischPRL15} oriented by an external magnetic field. The total permanent magnetic dipole moment of the molecule is taken to be equal to twice that of a single atom. The van der Waals coefficient for the collision between two Er$_2$ molecules is taken equal to four times the van der Waals coefficient between two Er atoms in their ground level.  

For the examples with the strongest permanent dipole moments $d$ in
Table~\ref{tab:dip-dip}, the characteristic distance $D$ is huge and
the energy $E_D$ is very small. For instance, the temperature
associated to $E_D$ varies from the nano-kelvin range for KRb down to
the femto-kelvin range for LiCs. Simultaneously the spatial range
increases from a few micro-meter up to a few milli-meter. For magnetic
atoms interacting via  dipole-dipole interaction, the interaction
length is smaller, a few hundred nano-meter, corresponding to much
higher temperatures, from $\sim 0.1\,\mu$K for Dy up to  $\sim
0.1\,$mK for Cr, whereas for molecular partners such as Er$_2$ it
corresponds to micro-kelvin.

For permanent dipoles, in order to compare the strength of the dipole-dipole interaction to the strength of the non-resonant light interaction, we introduce the critical laser intensity $\mathcal I_c$ for which the two become equal. 
It is important to note that in reduced units the value  of the critical intensity does not depend on the polarizabilities,
\begin{equation}\label{eq:Ic}
{\mathcal I}_c=\frac{3\,(2 \mu)^{3/4}}{4 \pi \epsilon_0 \hbar^{3/2} C_6^{1/4}} d_1 d_2\,.
\end{equation}
For collisions between aligned polar molecules, the critical intensity $\mathcal I_c$ is rather large, equal to 140, 570 and 3440$\,$ru for KRb, RbCs and LiCs respectively, see Table~\ref{tab:dip-dip}. In contrast, for collisions of magnetic atoms, the critical intensity is much smaller, equal to 1.5, 3.9 and 7.7$\,$ru  for Cr, Er, Dy atoms. Magnetic molecules represent an intermediate case, with $\mathcal I_c$=$19\,$ru for collisions between Er$_2$ molecules. These differences in $\mathcal I_c$ reflect the fact that the strength of the magnetic dipole-dipole interaction between highly magnetic atoms, transition metals and lanthanides, is smaller than the strength of the electric dipole-dipole interaction between molecules with large permanent electric dipole moment. 

With a number of good candidates at hand, we proceed to analyze non-resonant light control of the scattering volume. To this end, we need to account for how the particles are trapped.

%+++++++++++++++++++++++++++++++++++++++++++++++++++++++++++++++++++++++++++++++
\section{Control of  $p$-wave collisions}
\label{sec:predic-v}
%+++++++++++++++++++++++++++++++++++++++++++++++++++++++++++++++++++++++++++++++
%
When atoms or molecules are confined in a magneto-optical trap (MOT)
with an extension of up to a few millimeters, the confinement is very
weak and the interparticle distance can be considered to extend to
infinity. It is then possible to approximately assume the collision
partners to freely move in space. In this case, the asymptotic model
with universal nodal lines can be used to determine the
intensity-dependence of the generalized scattering volume
$v_m(\mathcal I,x_{00})\equiv \mathcal M^0_{BC2}$, as described in
Paper I~\cite{paperI}. This will be done in Sec.~\ref{subsec:weak},
where we pay particular attention to identifying intensities for which
a bound state lies at the dissocation limit and the generalized scattering volume diverges.

For strong confinement, as realized in an optical dipole trap or in
optical lattices, it is no longer possible to consider cold collisions
in free space. We examine, in Sec.~\ref{subsec:traps}, the case where
the characteristic length of the trap (assumed to be isotropic and
harmonic) is larger than $\sigma$, limiting the values of the
non-resonant field intensity to relatively small values, so that the
equivalent dipole length 
$D=\mathcal I\,\sigma /6$,  Eq.~\eqref{eq:D-sigma}, 
remains smaller than the characteristic trap length. We adapt the asymptotic model with universal nodal lines to the calculation of the trap energy levels in the presence of both dipole-dipole and short-range interactions. 

Finally, in Sec.~\ref{subsec:traps-x00}, we show the close connection between cold collisions in free space and in an isotropic harmonic trap. To this end, we  relate the energy shift of the $\ell$=1 trap levels to the generalized $p$-wave scattering volume.
%----------------------------------------------------------------------------------
\subsection{Free particles or weak confinement}
\label{subsec:weak}
%----------------------------------------------------------------------------------
%
%%%%%%%%%%%%%%%%%%%%%%%%%%%%%% figure contours %%%%%%%%%%%%%%%%%%%%%%%%%%%%%%%%%%%%%%
\begin{figure}[tb]
  \centering
  \includegraphics[width=.99\linewidth]{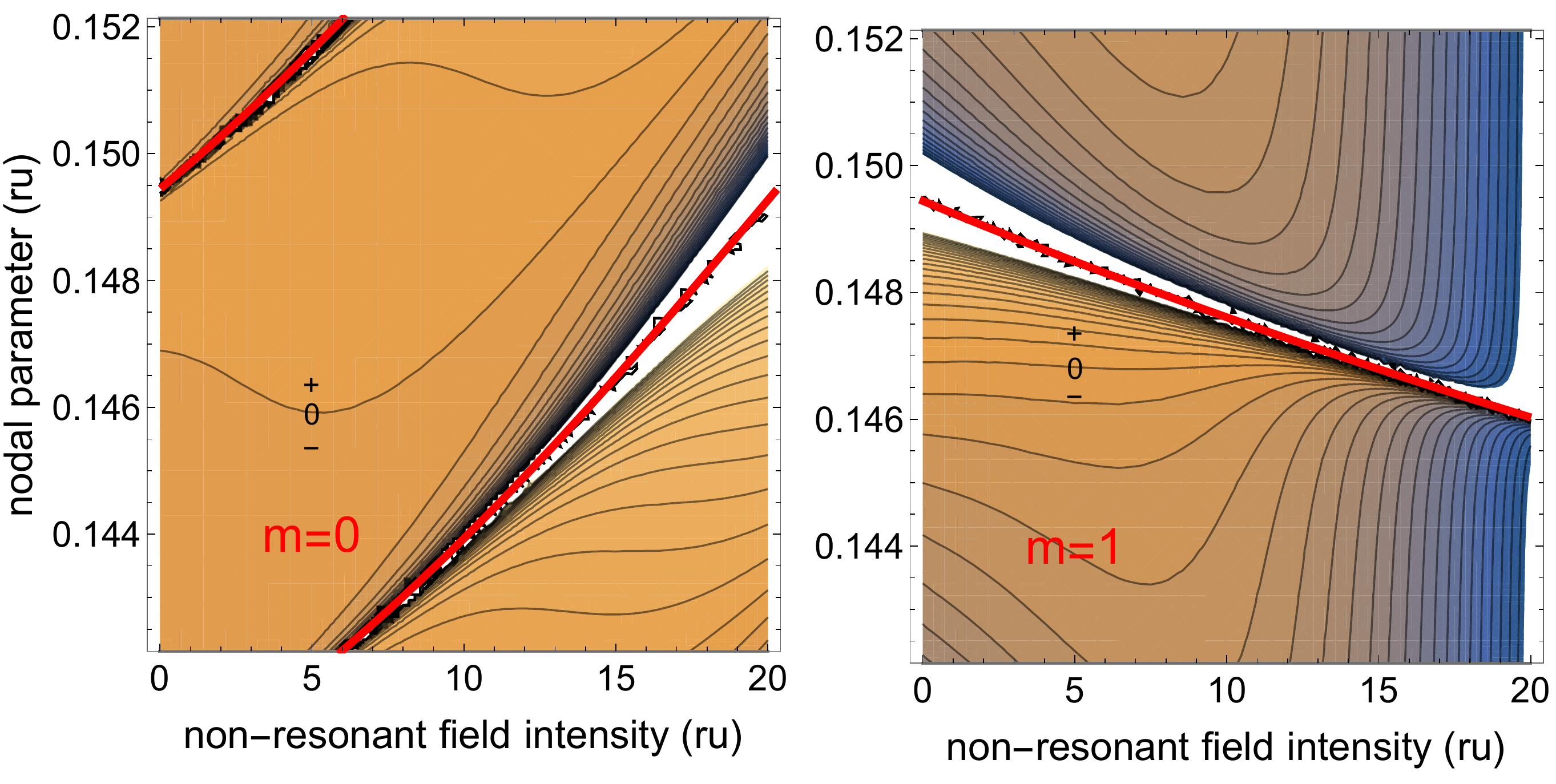}
  \caption{Generalized scattering volume as a function of the non-resonant field
    intensity $\mathcal I$ and the nodal parameter $x_{00}$ for $m=0$
    (left) and $m=1$ (right) in a single channel calculation
    ($\ell=1$). The gray contour lines vary in 40 steps between -150
    and 150 for $m=0$ and -2 and 2 for $m=1$ with the zero line labelled and the plus and minus signs indicating the regions of positive and negative values of the scattering volume. 
		The red thick lines
    indicate the values  of $\mathcal I$ and $x_{00}$ for which the field-dressed scattering volume diverges.
    Note that the absolute value of the scattering volume is in general 
    much larger for $m$=0 than for $|m|=1$, and that the width of the divergences is  generally also much larger, increasing with the light intensity for $m=0$, 
    while decreasing for $|m|=1$.
  \label{fig:SV_contour}
}
\end{figure}
%%%%%%%%%%%%%%%%%%%%%%%%%%%%%%%%%%%%%%%%%%%%%%%%%%%%%%%%%%%%%%%%%%%%%%%%%%%%%%%%%%%%%%%
% 
We first consider confinement of the colliding particles that is so weak that it can, to a good approximation, be neglected altogether.
When using the asymptotic model with universal nodal lines, a given
pair of colliding atoms  is characterized by its field-free $s$-wave
scattering length $a$ or, equivalently, by the nodal parameter
$x_{00}$, i.e., a node position of the corresponding field-free
$s$-wave threshold wave function~\cite{LondonoPRA10}. 
This approach is easily generalized to account for the presence of
non-resonant light with intensity $\mathcal I$ for both $s$-wave~\cite{CrubellierNJP15a} and
$p$-wave collisions, cf. Paper I~\cite{paperI}, where $a$, respectively $x_{00}$,
determines the colliding species. 
A general picture of the behavior of the scattering volume as a
function of the non-resonant light intensity for all pairs of particles is thus
obtained in terms of a contour plot, as shown in
Fig.~\ref{fig:SV_contour} for a single-channel calculation with
$\ell=1$. The range chosen for $x_{00}$ corresponds to one
quasi-period of the field-free $s$-wave scattering length varying from
$-\infty$ to $+\infty$, cf. Paper~I~\cite{paperI}.
Two singularities are observed in Fig.~\ref{fig:SV_contour}
for $m=0$, and one for $|m|=1$. These are indicated by the thick red
lines and correspond to infinitely strong interactions between the
colliding particles. For $m=0$ and a nodal parameter $0.1495\le x_{00}\le
0.1505$ (corresponding to a field-free $s$-wave scattering length in the range $0.9724-1.414$~ru), less than about 2$\,$ru or $2\,$GW/cm$^2$ of non-resonant light intensity is sufficient to effectuate a huge change of the
generalized $p$-wave scattering volume. Such an $s$-wave scattering
length is found for a mixture of $^7$Li and $^{40}$K, colliding in the
lowest triplet state.
Similarly, for $|m|=1$, the lowest intensities to realize a divergence of the scattering volume are needed for species characterized by a nodal parameter
$0.1490\le x_{00}\le 0.1495$ or, resp., a field-free $s$-wave
scattering length in the range $0.8428-0.9724$~ru, such as the
interspecies triplet scattering length of  $^{41}$K and
$^{87}$Rb~\footnote{The values of the scattering lengths are taken from
  Table I of Ref.~\cite{LondonoPRA10}}. 

%
%%%%%%%%%%%%%%%%%%%%%%%%%%%%%%% figure any dimer %%%%%%%%%%%%%%%%%%%%%%%%%
\begin{figure*}[tb]
  \centering
  \includegraphics[width=.99\linewidth]{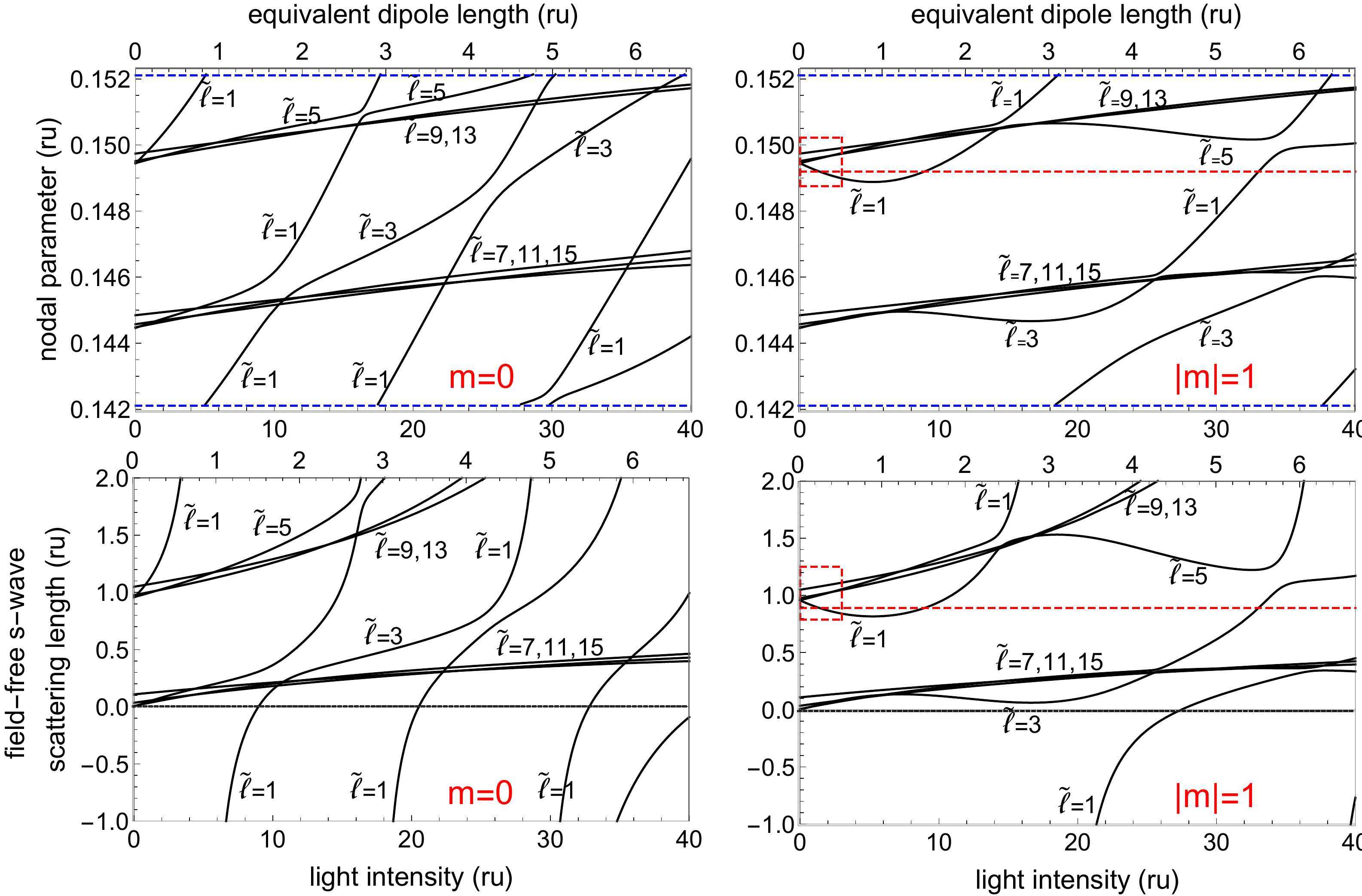}
  \caption{Map of the singularities of the generalized scattering volume
    (analogous to the red lines in Fig.~\ref{fig:SV_contour})
    as a function of nodal parameter $x_{00}$ (top panels),
    respectively the field-free s-wave scattering length $a$
    (bottom panels) and the non-resonant light intensity for $m$=0
    (left) and $|m|$=1 (right). The scaling with the equivalent
    dipole length, $D$ in Eq.~\eqref{eq:D-sigma}, for the
    anisotropic  interaction is also shown
    in the top horizontal axis in each panel. The singularities are
    obtained in terms of the appearance of a zero energy bound
    state and correspond to infinitely strong interaction
    between the particles.
    The calculations have been performed for odd $\ell$-values from 1 to
    17 ($n$=9 channels). 
    The horizontal blue dashed lines in the top graphs indicate the values of the nodal parameter for which the field-free $s$-wave scattering length is infinite. 
The curves corresponding to $\widetilde{\ell}$=5,9,13 (resp. $\widetilde{\ell}$=7,11,15) are grouped in a roughly horizontal beam starting from an initial value of approximately  $x_{00}\sim0.15$ (resp. $x_{00}\sim0.145$) in the top graphs and from an initial value $a\sim1$  (resp. $a\sim0$) in the bottom ones. The data within the red box are shown in more detail in Fig.~\ref{fig:inset}, and the red dashed lines indicate the cases that will be examined in Figs.~\ref{fig:i-variable} and \ref{fig:i-variable-deux}, with the red box corresponding to the bottom part of Fig.~\ref{fig:i-variable}. }
  \label{fig:scatt-any}
\end{figure*}
%%%%%%%%%%%%%%%%%%%%%%%%%%%%%%%%%%%%%%%%%%%%%%%%%%%%%%%%%%%%%%%%%%%%%%%%%%%%%%
%
The picture in Fig.~\ref{fig:SV_contour} is only of illustrative character due to the single channel approximation. A more quantitative picture is obtained in multi-channel calculations. Figure~\ref{fig:scatt-any} shows, for $n=9$ coupled channels, the singularities of the generalized scattering volume as a function of the non-resonant light intensity and the field-free $s$-wave scattering length (bottom), respectively the nodal parameter (top). The left and right-hand sides  of Fig.~\ref{fig:scatt-any}  correspond to  $m$=0 and $|m|$=1, respectively.   
For simplicity, only the singularities (corresponding to the red thick lines in Fig.~\ref{fig:SV_contour}) are shown and the contours are omitted. The bound states, whose occurrence at threshold causes the singularity, are labelled by $\widetilde \ell$, in reference to the $\ell$-channel with the largest weight in the field-dressed wave function. 
For the lowest two values, $\widetilde{\ell}$=1 and $\widetilde{\ell}$=3, 
the singularity curves vary rapidly and almost linearly as
a function of the non-resonant light intensity ${\mathcal I}$,
especially for $\widetilde \ell$=1 and $m$=0 (left part of Fig.~\ref{fig:scatt-any}). In this case, the
occurence of a bound level at threshold depends only to a limited
extent on the $s$-wave scattering length. Rather, it is  essentially
determined by the non-resonant field intensity, i.e., the anisotropic
long-range interaction. For $\ell=1$, $|m|=1$ (top right part of Fig.~\ref{fig:scatt-any}), a negative slope of the singularity curve is observed at low intensity. This is caused by the repulsive character of the effective adiabatic potential, cf. Table II in Paper I~\cite{paperI}. For higher intensity, the coupling with the other channels becomes dominant, turning the slope of the singularity curve positive, as for all the other $(\ell, m)$ values. 

\begin{figure}[tb]
  \centering
  \includegraphics[width=0.99\linewidth]{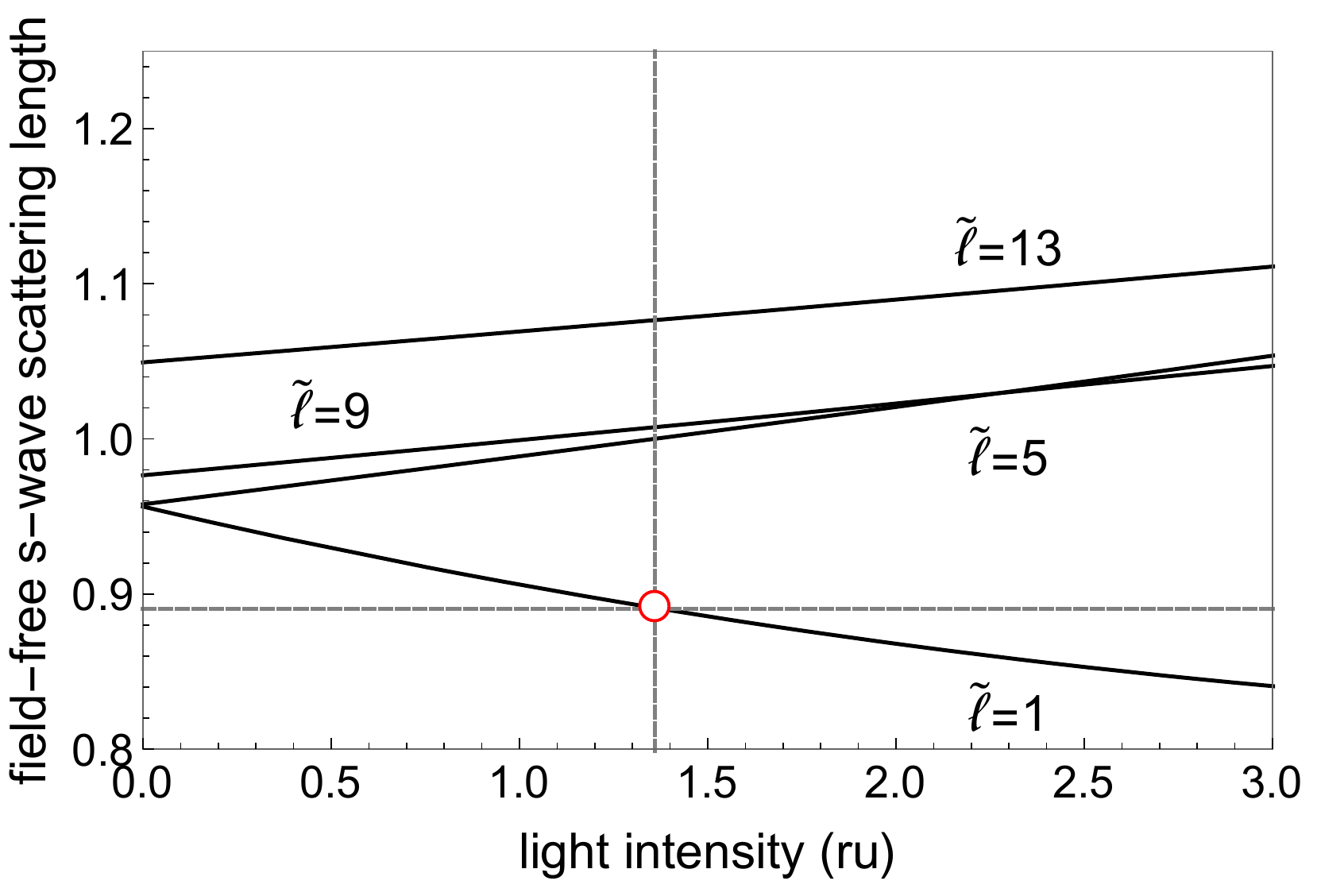}
  \caption{\label{fig:inset} 
    Inset of the bottom  right part of Fig.~\ref{fig:scatt-any}, i.e.,
    map of the singularities of the generalized scattering volume as a
    function of the field-free $s$-wave scattering length and
    non-resonant light intensity for $|m|=1$. Horizontal gray line:
    field-free $s$-wave scattering length equal to 0.891$\,$ru
    (corresponding to $x_{00}=0.1492\,$ru), the value  used 
    Figs.~\ref{fig:i-variable} and~\ref{fig:i-variable-deux}; vertical gray line: predicted value of
    the position of the singularity (the red circle simply marks their
    intersection). 
	}
\end{figure}
For the larger values of $\widetilde{\ell}$, singularities appear for
a field-free $s$-wave scattering length approximately equal to zero
(or, equivalently, $x_{00}$=0.149481~ru), for $\widetilde \ell=5$, 9, 13, 
and approximately equal to 0.96~ru (resp., $x_{00}$=0.144652~ru) 
for $\widetilde \ell =7$, 11, 15~\footnote{Note that it is necessary
  to introduce the channel $\ell_{max}$ (here $\ell_{max}=17$) to
  obtain converged results for the channels $\ell \le \ell_{max-2}$.}. 
These two values of the field-free
$s$-wave scattering length are close to those predicted by the
analytical model of Gao~\cite{GaoEPJD04}. The 
corresponding singularity curves of the $p$-wave scattering volume vary
only slowly with the light intensity. 
This indicates that the corresponding field-dressed wave functions strongly depend on the short-range interaction and almost not 
on the anisotropic long-rang interaction. It can be understood in terms of 
the height of the rotational barrier which,
being proportional to $\propto \ell^3$, increases rapidly with 
$\ell$ but is barely modified by the non-resonant light at the studied
intensities. 

It is worth mentioning that the width of the singularity as a function
of intensity is not independent of the width as a function
$x_{00}$. Let us define the width of a singular function of the type
$y(x)=w/(x-x_p)$ with a pole at $x=x_p$ by $w$. At a given point in
the  ($\mathcal I$,$x_{00}$)-plane, the width along the first axis is
proportional to the width along the other one, with the
proportionality factor being equal to the opposite of the slope of the singularity curve.
%
%-------------------------------------------------------------------------------
\subsection{Strong confinement}
\label{subsec:traps}
%-------------------------------------------------------------------------------
%
%
%%%%%%%%%%%%%%%%%%%%%%%%%% figure 4 energy vs intensity un %%%%%%%%%%%%%%%%%%%%%%
\begin{figure}[tb]
  \centering
  \includegraphics[width=0.99\linewidth]{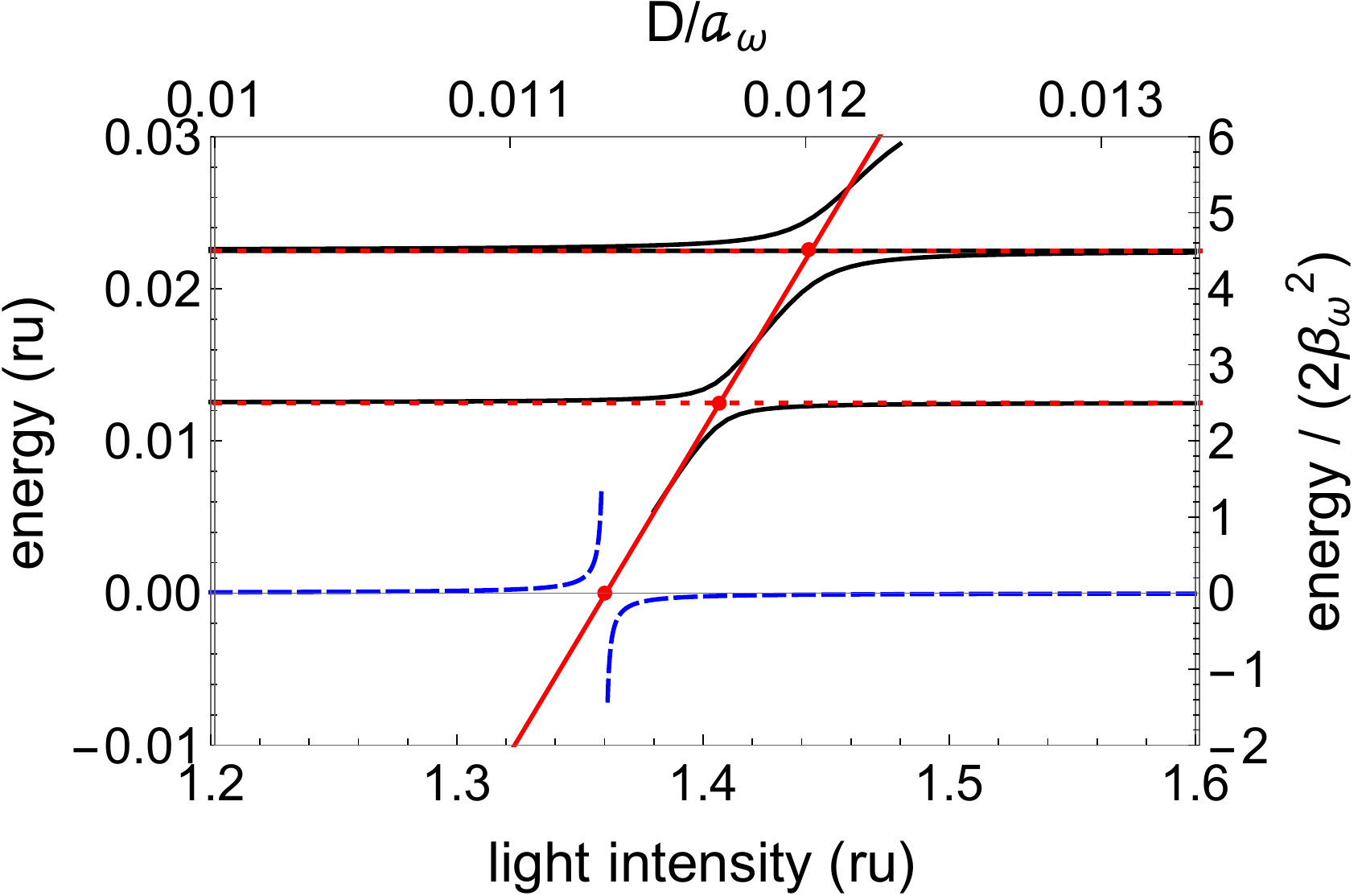}
  \caption{\label{fig:i-variable} 
Trap state energy as a function of the non-resonant light intensity
${\mathcal I}$ (black lines), compared to the energy of the field-free
trap states (red dashed lines), identified in the text.
The scales are reduced units (ru) of the van der Waals interaction
(left), reduced units (ru$(\omega)$) of the harmonic oscillator
(right), non-resonant light intensity (bottom), and ratio of the
equivalent dipole length $D$, Eq.~\eqref{eq:D-sigma}, to harmonic
oscillator length $a_{\omega}$ (top). 
The light-induced anticrossings 
of the field-free $\widetilde{\ell}$=$1$ trap states with the
intensity-dependent last bound $p$-state (for negative energies),
resp. $p$-wave shape resonance (for positive energies) for untrapped particles (red solid line) are clearly observable.
The calculation is performed for a three channel model with $\ell$=1,
3, 5, $|m|$=1 with $x_{00}$=0.1492~ru,  i.e.,  a field-free $s$-wave
scattering length of $0.891\,$ru. This intensity range corresponds to the first intersection of the red dashed horizontal line in Fig.~\ref{fig:scatt-any} 
with the black $\widetilde{\ell}$=1 curve.
The  blue dashed line displays the intensity dependence of 
the generalized scattering volume, calculated under the same
conditions ($\ell$=1, 3, 5, $|m|$=1 and $x_{00}$=0.1492~ru) and
multiplied for convenience by the factor $B$ (for $N$=0) of
Table~\ref{tab:trap} in Appendix \ref{sec:app}. 
	}
\end{figure}
%%%%%%%%%%%%%%%%%%%%%%%%%%%%%%%%%%%%%%%%%%%%%%%%%%%%%%%%%%%%%%%%%%%%%%%%%%%%%%%%%%
%
%
%%%%%%%%%%%%%%%%%%%%%%%%%% figure 5 energy vs intensity deux %%%%%%%%%%%%%%%%%%%%%
\begin{figure}[tb]
  \centering
  \includegraphics[width=0.99\linewidth]{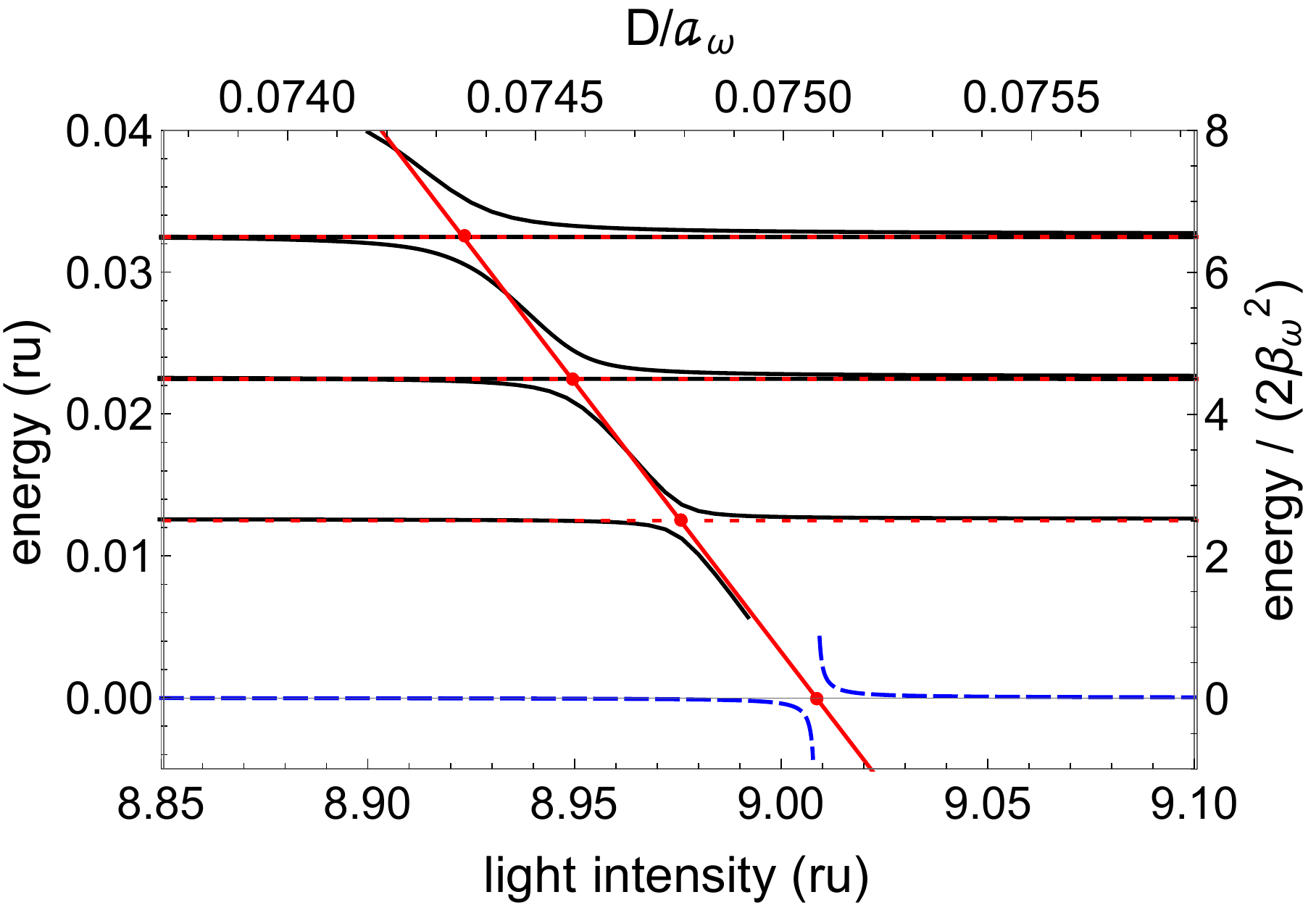}
  \caption{\label{fig:i-variable-deux} Same as Fig.~\ref{fig:i-variable} but for the second intersection of the red dashed horizontal line (at $x_{00}=0.1492\,$ru) in Fig.~\ref{fig:scatt-any}
    with the black  $\widetilde{\ell}=1$ curve.
  }
\end{figure}
%%%%%%%%%%%%%%%%%%%%%%%%%%%%%%%%%%%%%%%%%%%%%%%%%%%%%%%%%%%%%%%%%%%%%%%%%%%%%%%%%%
%
If the pair of particles is confined in an isotropic 3D harmonic
potential of frequency $\omega$, a term $\beta_{\omega} ^4 x^2$ has to
be added in the equation~\eqref{eq:asy}, describing the relative motion
in van der Waals reduced units, with
\begin{equation}
\label{eq:ru-osc-x}
\beta_{\omega}=\sigma \sqrt{\frac{\mu \omega}{\hbar}}=\sigma/a_\omega\, , 
\end{equation}
where $a_\omega$, the trap reduced unit of length, is related to the trap reduced unit of energy $\epsilon_\omega$ 
\begin{equation}
\label{eq:ru-osc-e}
\epsilon_\omega=\hbar \omega=\frac{\hbar^2}{\mu (a_\omega)^2}=2 \epsilon (\beta_\omega)^2 \,.
\end{equation}
With the unit factors $a_\omega$ and
$\epsilon_\omega$, the length $x_\omega$ (resp. energy $e_\omega$)
expressed in reduced units of the harmonic oscillator (ru($\omega$))
is related to the corresponding value $x$ (resp. $\mathcal E$) in van
der Waals reduced units (ru) by $x_\omega=x\beta_\omega$  ($e_\omega={\mathcal E}/(2\beta_\omega^2)$).
When strongly confined in a trap, where at large distance the trapping
potential $\propto x^2$ prevails, the particles will explore only a
limited range of the dipole-dipole interaction potential. We thus may
expect that, in the lowest positive-energy states of the trap, the
behavior of the inter-particle interaction will be close to the one
described by small $kx$ or, equivalently, by the threshold case,
$k$=0, cf. Sec.~II~B of Paper I~\cite{paperI}, where $k$ denotes the
wave number in van der Waals reduced units ($\mathcal E=k^2$).
Due to the trap, the spectrum possesses bound states only. The study
of the asymptotic phase shift of the scattering wave functions is thus
replaced by  analysis of the bound state energy shift with respect to
the energy of the unperturbed trap states in harmonic oscillator
reduced units, 
$e_\omega(N\ell)= (2N +\ell +3/2)\,$ru$(\omega)$ ($N\ge0$ integer). 
We consider here only the case where $a_\omega$ is larger than $\sigma$ and we limit the non-resonant light intensity to relatively small values, so that the characteristic length of the dipole-dipole interaction remains always much smaller than $a_\omega$. 

To calculate the energy ${\mathcal E}$ of the bound states, we adapt the general procedure described in Ref.~\cite{LondonoPRA10}. The wave function satisfies boundary conditions both on the nodal lines and  at large distance $x_{max}$. The initial condition for the inward integration of the particular solution ${\bf f}^\ell({\mathcal E},x)$ is given by $f^\ell_{\ell'}({\mathcal E},x_{max})=\delta_{\ell,\ell'} \exp(-\beta_\omega^2 x_{max}^2/2)$ in the channel $\ell'$. 
Writing the physical wave function ${\bf z}({\mathcal E},x)$ as a linear combination of the solutions ${\bf f}^\ell({\mathcal E},x)$  and requiring it to vanish on the nodal lines yields the quantization condition for the energy.

The intensity dependence of the bound state energies, calculated with three odd $\ell$-values, $|m|$=1, and a trap potential $\beta_\omega$=0.05, i.e., a trap length $a_\omega$=$20\,\sigma$ %, for 
is studied in two different intensity ranges.
In both cases, the chosen nodal parameter is $x_{00}$=0.1492~ru and the field-free $s$-wave scattering length is equal to $0.891$~ru. 
For this choice of parameters, an untrapped pair of particles subject
to non-resonant light possesses two times a bound state with
${\widetilde \ell}$=1,  $|m|$=1 at threshold, as shown in
Fig.~\ref{fig:scatt-any},  for a light intensity $\mathcal I$ equal to
1.36~ru and 9.01~ru. The corresponding equivalent dipole lengths $D$,
cf. Eq.~\eqref{eq:D-sigma},  amount to  $0.23\,\sigma$ and $1.50\,\sigma$, as shown in Fig.~\ref{fig:scatt-any}.

The intensity regions around $\mathcal I=1.36\,$ru and 9.01~ru are explored separately in Figs.~\ref{fig:i-variable} and \ref{fig:i-variable-deux}. 
The relevant field-free trap states correspond to $N$=0, $\ell$=1 (the
lowest odd-$\ell$ trap level, with $e_\omega=5/2\,$ru($\omega$)),
$N=0$, $\ell=3$ and $N=1$, $\ell=1$ (the doublet of trap levels with
$e_\omega=9/2\,$ru($\omega$)) in Fig.~\ref{fig:i-variable} and, 
in addition, the triplet of trap levels with
$e_\omega=13/2\,$ru($\omega$), 
$N=0$, $\ell=5$, $N=1$, $\ell=3$ and $N=2$, $\ell=1$ in
Fig.~\ref{fig:i-variable-deux}. 

Two avoided crossings are observed in  Fig.~\ref{fig:i-variable}, around $\mathcal I=1.4$ and 1.43~ru. They are due to the strong coupling that the anisotropic interaction induces between each of the two $\widetilde\ell=1$ trap states (with $N=0$ and $N=1$) and the untrapped last bound  $\ell=1$-state together with its continuation as a $\widetilde\ell=1$ shape resonance. The $N=0$, $\ell=3$ trap state is not noticeably perturbed, see the essentially horizontal black line in Fig.~\ref{fig:i-variable}.
The red curve displaying the intensity dependence of the energy of the
last bound $p$-level (for negative energy) or the $p$-wave  shape resonance (for positive energy) for the untrapped pair crosses the red dashed curves representing the field-free $\ell=1$
trap states at the position of the anticrossings. In addition, the red
curve crosses the zero energy for the intensity at which the
generalized scattering volume diverges. 
The increase of the trap state energy with ${\mathcal I}$ in  Fig.~\ref{fig:i-variable} 
is related to the repulsive character of the anisotropic interaction
in the $\ell$=1, $|m|$=1 channel, as discussed in the previous
subsection and visible in the negative slope of the ${\widetilde
  \ell}$=1 curve at small $\mathcal I$ in Figs.~\ref{fig:scatt-any} and~\ref{fig:inset}.

In Fig.~\ref{fig:i-variable-deux}, the situation is similar to that in Fig.~\ref{fig:i-variable}, but the repulsive character of the dipolar interaction in the $\ell$=1, $|m|=1$ channel is superseded by the coupling with the other channels. As a consequence, and as is most generally the case, close to the divergence of the generalized scattering volume, the trap state energy  decreases with intensity in Fig.~\ref{fig:i-variable-deux}, 
and the slope of the singularity curve with  $\widetilde{\ell}$=1 in Fig.~\ref{fig:scatt-any} is positive near ${\mathcal I}=9$. 
Note that while the $N=1$, $\ell=3$ and $N=0$, $\ell=5$ trap states 
(black solid lines showing avoided crossings in Fig.~\ref{fig:i-variable-deux})
are strongly mixed together in the vicinity of the divergence of the
scattering volume, they are not noticeably mixed with the $N=2$,
$\ell=1$ state (horizontal black lines at $e_\omega=13/2\,$ru($\omega$) in Fig.~\ref{fig:i-variable-deux}).
The intensity dependence of the trap state energies in Fig.~\ref{fig:i-variable-deux} is similar to the dependence 
of the energy of two aligned identical bosonic dipoles under external
confinement with strength characterized by 
$D/a_\omega$~\cite{Kanjilal07}, where $D$ is the equivalent dipole length of Eq.~\eqref{eq:D}. Figure~2 of Ref.~\cite{Kanjilal07} shows the energy of lowest trapped level of the pair to dive down to negative energy close to the $D/a_\omega$ value at which the two-body potential supports a new bound state. Moreover, the same behavior is also predicted for identical fermions undergoing $p$-wave collisions \cite{Kanjilal04}. 

Our calculations suggest that it should be possible to control the formation of molecular bound states in $p$-wave collisions by non-resonant light. This would analogous to using non-resonant light to  create molecules from bosonic atoms in $s$-wave collisions, as discussed in Ref.~\cite{GonzalezPRA12}:
Slowly increasing the light intensity around $\mathcal I$=9.01~ru would transfer the particle pair at resonance from the lowest trap state to the last molecular bound state. The same is true for the situation depicted in  Fig.~\ref{fig:i-variable}, except that the intensity has to be decreased around $\mathcal I$=1.36~ru. The inverse process consisting of climbing the trap ladder upward by a rapid variation of the light intensity would also be possible. Making molecules with non-resonant light this way would be the generalization of an experiment carried out for $s$-wave collisions, in the vicinity of a Feshbach resonance, of fermionic $^{40}$K atoms in various hyperfine states confined in an optical 3D lattice~\cite{StoferlePRL06}. Note that the experimental results of \cite{StoferlePRL06} are reproduced by a model, also used below, describing the short-range interaction by a pseudo-potential with a scattering length independent of  energy \cite{Busch98,Bolda02,Kanjilal04}. Moreover, just as the scattering behavior discussed above can be tuned with either non-resonant light or dipole interaction strength, the 
formation of molecules also has its analogue for colliding dipoles. Specifically, 
the simultaneous variation of the lowest trap state energy (black curve) and  the generalized scattering volume (blue dashed curve) with the non-resonant light intensity in Fig.~\ref{fig:i-variable-deux} is reminiscent of the dependence of energy and scattering parameter on dipole coupling strength 
in Refs.~\cite{BortolottiPRL06,RonenPRA06}. In both cases, the effective interaction is increasingly attractive and the scattering parameter negative at the left of the resonance. Conversely, the interaction is decreasingly repulsive and the scattering parameter positive to the right of the resonance. In both cases, at resonance, the pair is transferred  from the lowest trap state to the last molecular bound state. The main difference lies in the existence, in the case of non-resonant light control, of a small shift of the pole of the generalized scattering volume relative to the position of the lowest anticrossing of the trap energies. This shift comes from the finite slope of the energy as a function of intensity, cf. the red curve in  Fig.~\ref{fig:i-variable-deux}, which is due to the presence of the van der Waals potential and the energy-, $\ell$- and intensity-dependence of the nodal lines.

%-------------------------------------------------------------------------------
\subsection{Connecting $p$-wave scattering control with non-resonant light in weak and strong confinement}
\label{subsec:traps-x00}
%
%
%%%%%%%%%%%%%%%%%%%%%%%%%% figure 6 trap m=0 %%%%%%%%%%%%%%%%%%%%%%%%%%%%%%%%%%%%%
\begin{figure}[tb]
  \centering
  \includegraphics[width=0.99\linewidth]{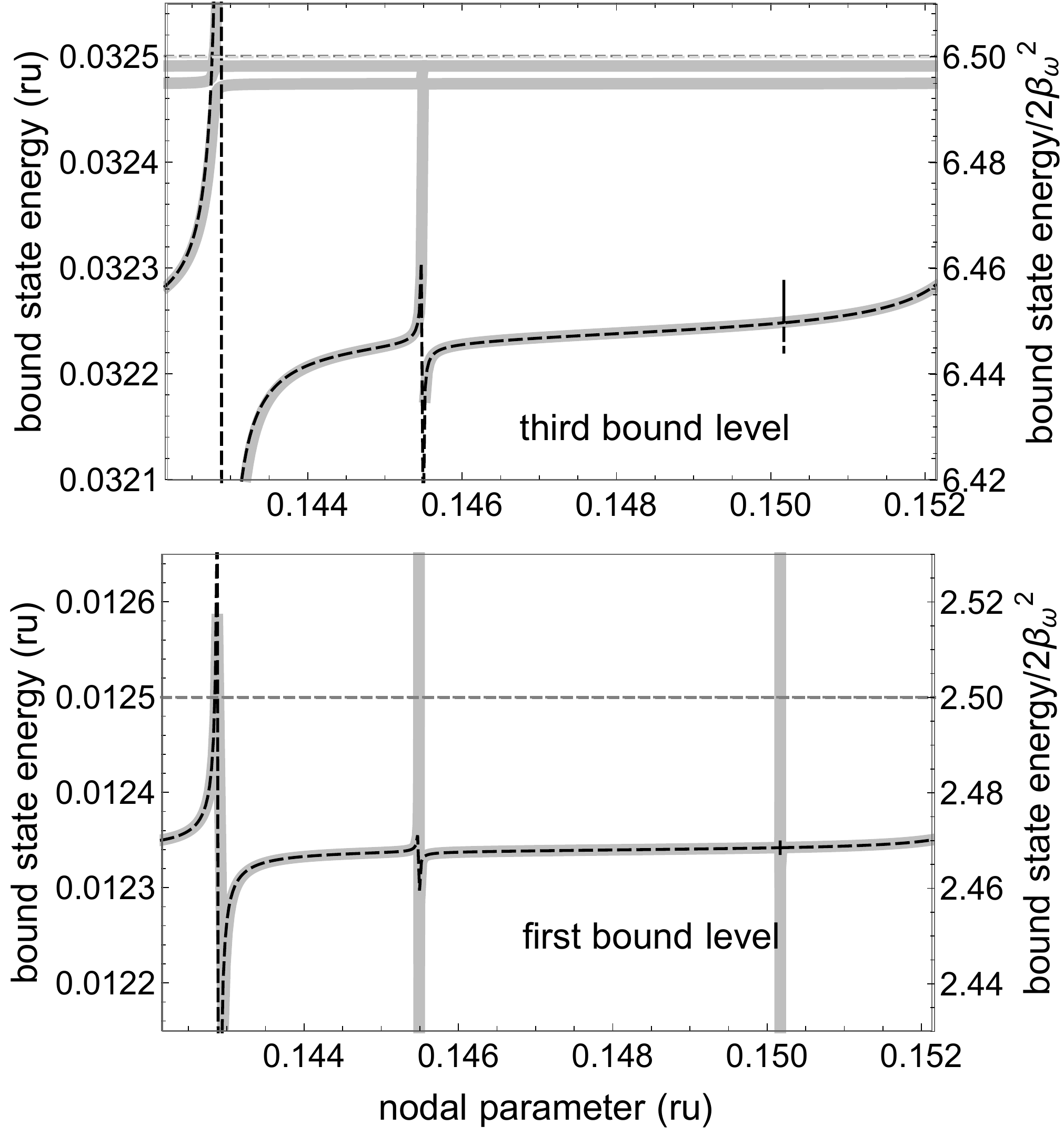}
  \caption{\label{fig:trap-0} Trap state energy (thick gray curves) of
    $m$=0 states as a function of the nodal parameter $x_{00}$ for
    ${\mathcal I}$=6~ru and $\beta_{\omega} $=0.05, in van der Waals
    reduced units on the left-hand side and in harmonic oscillator
    reduced units on the right-hand side for the lowest level of the
    trap with $N$=0, $\ell$=1 (bottom) and the third level, i.e., the
    set of states with $2N+\ell$=5, whose three-fold degeneracy is
    removed by the dipolar interaction (top). The black dashed lines
    display the $x_{00}$-dependence of the field-dressed generalized
    scattering volume ${\mathcal M}_0$ of untrapped particles,
    multiplied by $B=1.355 10^{-6}$~ru ($B=5.9 10^{-6}$~ru) and
    vertically shifted by 0.01233850~ru (0.03223315~ru), to fit the
    scale of the figure. This corresponds to a global shift of  $A=-0.0001615$~ru ($A=-0.00026685$~ru) of the first (third) trap level. The interaction-free trap levels are indicated by the thin dashed gray lines. 
The calculations are performed in a three-channel model ($\ell$=1, 3, 5). 
	}
\end{figure}
%%%%%%%%%%%%%%%%%%%%%%%%%%%%%%%%%%%%%%%%%%%%%%%%%%%%%%%%%%%%%%%%%%%%%%%%%%%%%%%%%%%
%
%
%%%%%%%%%%%%%%%%%%%%%%%%%% figure 7 trap m=1 %%%%%%%%%%%%%%%%%%%%%%%%%%%%%%%%%%%%%%
\begin{figure}[tb]
  \centering
  \includegraphics[width=0.99\linewidth]{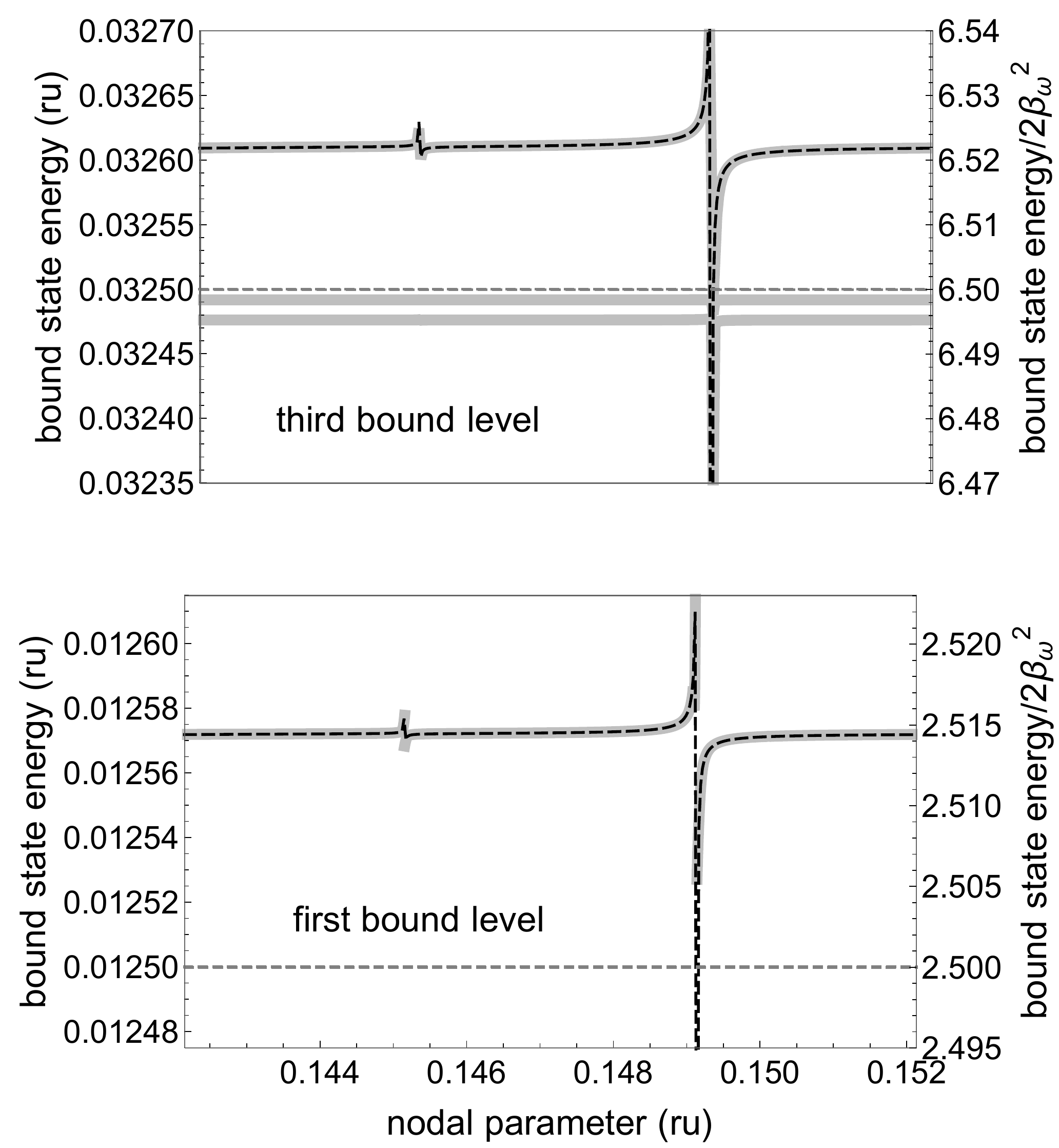}
  \caption{\label{fig:trap-1} Same as Fig~\ref{fig:trap-0}, but for
    $|m|$=1 states. 
The scaling parameters for  ${\mathcal M}_0$ are $B$=$1.5\,10^{-6}$~ru ($B$=$5.9\,10^{-6}$~ru) and shifts of $0.0125725$~ru ($0.032611$~ru), corresponding to a global shift of  $A=0.0000725$~ru (resp. $A=0.000111$~ru), of the first (third) level.
	}
\end{figure}
%%%%%%%%%%%%%%%%%%%%%%%%%%%%%%%%%%%%%%%%%%%%%%%%%%%%%%%%%%%%%%%%%%%%%%%%%%%%%%%%%%%%

For a pair of trapped particles subject to non-resonant light, it is
possible to connect the intensity dependence of the $p$-wave
scattering properties in the weak and strong confinement case by
studying the intensity dependence of either the field-dressed
generalized scattering volume $\mathcal M_0$ or the trap state energy
shifts.
As discussed in detail in Sec.~III in Paper I~\cite{paperI}, the
field-dressed generalized scattering volume, $\mathcal M_0$, depends
on the nodal parameter $x_{00}$ that is characteristic of the
short-range interactions in the field-free untrapped pair and displays
singularities, i.e., signatures of the appearance of $\widetilde \ell$-bound states  at threshold.
Moreover, for a given intensity, many singularities (for different $\widetilde\ell$) are found, cf. Fig.~\ref{fig:scatt-any}, whereas for a given nodal parameter, i.e., for a given choice of particles, intensity intervals that contain divergences have to be carefully selected.
So we first examine the $x_{00}$-dependence %behavior of the light- and interaction-induced 
of the trap state energy shifts for a fixed light intensity $\mathcal
I$.

Without any interaction in the pair, the trap energy for a state with quantum
numbers $N$, $\ell$=1, $m$ is given by
${\mathcal E}(N\ell m)$=$2(\beta_\omega)^2(2N+\ell+3/2)\,$ru in van
der Waals reduced units or, equivalently, $e_\omega(N\ell
m)$=$(\beta_\omega)^2(2N+\ell+3/2)\,$ru($\omega$), in harmonic
oscillator reduced units.
The interparticle
interaction in the untrapped pair induce a shift $\Delta {\mathcal E}_{N,\ell=1,m}$ that
will depend 
on the characteristics of this interaction, i.e., 
the short-range interactions  described by the nodal parameter $x_{00}$ (or equivalently the field-free scattering length),  the van der Waals asymptotic interaction
 $-1/x^6$ in reduced units and the anisotropic dipolar interaction induced by the light.

Figures~\ref{fig:trap-0} and~\ref{fig:trap-1} display $x_{00}$-dependence of the  energy of the first (or lowest) and third trap level for a non-resonant light intensity of  $\mathcal I$=6 ru ($D$=$\sigma$) and a trapping potential with $\beta_\omega=0.05$ ($a_\omega$=$20\sigma$). $m=0$ in Fig.~\ref{fig:trap-0} and $|m|=1$ in Fig.~\ref{fig:trap-1}.
As everywhere in this paper, the $x_{00}$ range is chosen so that the corresponding field-free $s$-wave scattering length varies once from $-\infty$ to $+\infty$. 
Note that the lowest trap level is non-degenerate, whereas the third one is triply degenerate. 
The short-range interaction of the particles and the coupling to the non-resonant light produce an $\ell$-dependent energy shift that removes the degeneracy. Therefore three separate grey curves are observed in the top of Figs.~\ref{fig:trap-0} and~\ref{fig:trap-1}.
In the lowest $\widetilde \ell=1$ adiabatic potential, $V_{\ell=1}(x)
\sim -c_3/x^3$, the 
interaction is attractive with $c_3>0$  for $m$=0 (resp. repulsive
with $c_3<0$ for $|m|$=1), and the trap state energy is shifted toward
lower (resp. higher) energy, cf. the difference between the thin
dashed gray lines -- corresponding to pure trap states -- and the
thick gray curves -- the perturbed trap states. This difference is essentially constant, except close to a divergence.
The adiabatic potentials $\widetilde \ell \ge 3$ are all attractive, resulting in a negative energy shift of the trapped states $\widetilde \ell$=3 and 5. The $\widetilde \ell$=1 trap states  show a $x_{00}$-dependence of their energy in the vicinity of the unperturbed trap level energy that is quite similar to that of the field-dressed generalized scattering volume ${\mathcal M}_0^m(x_{00})$. To visualize this in Figs.~\ref{fig:trap-0} and~\ref{fig:trap-1}, we have 
scaled the $x_{00}$-dependence of the field-dressed generalized scattering  volume shown in Fig.~1 of Paper I~\cite{paperI}
to fit the energy range of the trap levels and included the resulting curves with black dashed lines. 
The three resonances of  ${\mathcal M}_0^m(x_{00})$ observed in
Fig.~\ref{fig:trap-0} for $m$=0, which are  associated to the channels
$\ell$=1, 3 and 5,  appear at exactly the $x_{00}$-value as those of
the trap state energies. The same is true for the $|m|$=1 resonances in Fig.~\ref{fig:trap-1}, except that the resonance with $\ell$=5, which is extremely narrow, is not resolved in our calculations.

The ease with which the numerical results for the energies in the trapped case and the generalized scattering volume in the case of free collisions 
in  Figs.~\ref{fig:trap-0} and~\ref{fig:trap-1} can be connected
suggests a closer inspection of their relation. The
$x_{00}$-dependence of the field-dressed generalized scattering volume ${\mathcal M}_0^m(x_{00})$ was deduced in Sec.~III of Paper I~\cite{paperI}. 
Since it was only necessary to scale and shift ${\mathcal M}_0^m(x_{00})$ as a function of  $x_{00}$ in order to display it together with the trap state energies, the simple ansatz for the interaction-induced shift in energy,
\begin{equation}
\label{eq:osc-shift}
 \Delta {\mathcal E}_{N,\ell=1,m}(x_{00})=A + B {\mathcal M}_0^m(x_{00})\, ,
\end{equation}
should be sufficient. 
The linear equation~\eqref{eq:osc-shift} is written for BC2 reference
functions, cf. Table~I in Paper~\cite{paperI}, and assuming that
$\Delta {\mathcal E}_{N,\ell=1,m} \ll 4 (\beta_\omega)^2$, i.e., the
energy shift is much smaller than the level spacing of interaction-free trap levels.
In Eq.~\eqref{eq:osc-shift}, the parameters $A$ and $B$ depend on the
confinement $\beta_\omega$ and the quantum numbers $N$, $\ell$, and $m$, but are independent of the nodal parameter.
For the lowest trap level, i.e., $N$=0, $\ell$=1, it is possible to determine 
this dependence. To this end, we calculate the energy related to the
trap potential for a wave function, $f_{pert}(x)$, which, 
for $\beta_\omega x<<1$, has the same behavior as the untrapped threshold $p$-wave function $u(x)$.
While the trap wave function without any interaction is proportional to $ x^2 \exp(-(\beta_{\omega} x)^2/2)$, in the presence of interactions the ansatz
$f_{pert}(x)= u(x) \exp(-(\beta_{\omega} x)^2/2)=(x^2-ax-b-c \ln(x)/x -d/x-{\mathcal M}_{0}/x) \exp(-(\beta_{\omega} x)^2/2)$ ensures the correct behavior at long interparticle distances. In this ansatz,
$a$, $b$, $\ldots$ 
depend on the $m$- and ${\mathcal I}$-dependent parameters $c_3$,
$c_4$, $\ldots$, which describe the asymptotic form of the effective
potential for $p$ waves, cf. Table II of Paper I~\cite{paperI} while
$u(x)$ is reported in Eq.(B7) of Ref.~\cite{paperI}. 
With this ansatz, the mean potential energy becomes
\begin{equation}
 {V}_{pert}=\frac{\int \beta_{\omega} ^4 x^2 f_{pert}(x)^2 dx}{\int f_{pert}(x)^2 dx}\,,
\end{equation}
where the integration runs from a small value to $\infty$. Using the
virial theorem for the harmonic oscillator, the total energy is twice
this value. Comparing the total energy of the trapped interacting pair
to the trap level without interaction, we find, 
for the lowest trap level,
\begin{eqnarray}
\label{eq:AB-LK}
A=-\frac{4c_3}{3 \sqrt\pi} \beta_{\omega}^3\,,\quad 
B=\frac{8}{\sqrt\pi} \beta_{\omega}^5\,,
\end{eqnarray}
in good agreement with the numerical results, that were obtained for
$m$=0 and $m$=$\pm1$ in both single-channel ($\ell$=1) and
multi-channel ($\ell$=1, 3, 5) calculations. For instance, the
estimates of $A$ and $B$ for the lowest level shown in
Fig.~\ref{fig:trap-0} are $1.41\,10^{-6}$~ru and -0.0001504~ru, to be
compared with the values quoted in the caption of the figure, i.e.,
$1.355\,10^{-6}$~ru and -0.0001615~ru. Similarly, for $|m|$=1, i.e.,
Fig.~\ref{fig:trap-1}, the estimates for $A$ and $B$ are
$1.41\,10^{-6}$~ru and 0.00007523~ru, to be compared to
$1.5\,10^{-6}$~ru and 0.0000725~ru. However, this method is not
suitable to determine $A$ and $B$ for higher trap levels, since the
ansatz $f_{pert}(x)$ is built upon the (zero-energy) threshold wave
function $u(x)$. We therefore resort to a more general procedure in
Appendix \ref{sec:app} where we find $A$ and $B$ to be determined by
the long-range dipolar 
and short-range parts of the interactions, respectively. This is not surprising since $\mathcal M^m_0(x_{00})$ (which multiplies $B$) unambiguously characterizes the contribution of the short-range interactions to ultracold dipolar scattering, as discussed in detail in Paper I~\cite{paperI}. It is the important role of the short-range interactions that also explains the close connection between the cases of weak (or no) and strong confinement.

%+++++++++++++++++++++++++++++++++++++++++++++++++++++++++++++++++++++++++++++++
\section{Generalized scattering volume and orientation of the interparticle axis}
\label{sec:orient}
%+++++++++++++++++++++++++++++++++++++++++++++++++++++++++++++++++++++++++++++++
%
%
We  study in the following the interdependence of the generalized
scattering volume and the orientation of the interparticle axis
relative to the direction of the two dipoles, induced dipoles in the
case of non-resonant light control or permanent aligned dipoles in the case of
polar molecules. Remember that we assume the dipoles to be aligned
along the laboratory $Z$ axis, cf. Sec.~\ref{subsec:hamil}. 
Our focus is on the orientation  of the interparticle axis relative to
the direction of the  dipoles in the case where the non-resonant light
is used to induce a divergence of the generalized scattering volume. Nevertheless,
this still is formally identical to the case of permanent dipoles,
provided the direction is fixed.
While it is challenging to solve for the complete scattering dynamics, insight can already be gained by examining the orientation as a function of interparticle distance.

Due to the symmetry of the problem, the representation of the
Hamiltonian in the basis of the spherical harmonics is diagonal in $m$ and
depends on the absolute value of $m$ only. The eigenfunctions with $m$=0 and
$m=\pm1$ are independent solutions of the eigenvalue problem of two
different $m$-dependent Hamiltonians, denoted for simplicity by $H_0$
and $H_1$ and obtained from the 2D Hamiltonian
$H$~\eqref{eq:2D_Hamil}, 
\begin{equation}
  H^{\ell,\ell'}_m = \langle \,Y_{\ell,m} | H |Y_{\ell',m} \,\rangle\,.
\label{eq:H_m}
\end{equation}
In an experiment, it is impossible to select a given value of $m$ (except for very specific cases, such as samples in the shape of a pancake or a needle). Therefore, in general, 
the scattering states are a linear combination of two solutions, 
one with  $m=0$ and one with $m=\pm 1$.
At a specific interparticle distance, the ratio $\tan(\alpha)$ of the  $m=0$ and $m$=$\pm1$ coefficients fixes the orientation of the interparticle axis with respect to the direction of the two dipoles. 
This orientation will be a function of the interparticle distance, except if it is fixed by the geometry of the sample (in case of confinement to, e.g., a disk or a needle).

We distinguish below between freely movable and geometrically confined
dipoles. For freely movable dipoles, we first inspect a single channel
model in Sec.~\ref{subsec:orient} and generalize to the multi-channel
case in Sec.~\ref{subsec:orient-multi}. The situation of a particles
that are geometrically confined due to a specific shape of the trap 
is discussed in Sec.~\ref{subsec:orient-fixed}.

%ooooooooooooooooooooooooooooooooooooooooooooooooooooooooooooooooooooooooooooooo
\subsection{Orientation of the interparticle axis at short internuclear distance: single channel}
\label{subsec:orient}
%ooooooooooooooooooooooooooooooooooooooooooooooooooooooooooooooooooooooooooooooo
% 

We start with the single channel approximation (with $\ell=1$) because of its simplicity and in order to gain some first intuition.
Due to the symmetry around the laboratory fixed Z axis, it is sufficient to analyze the wave function in the Z-X plane ($\phi$=0). This reduces the angular part to its dependence on $\theta$, the angle between the interparticle axis and the laboratory fixed $Z$ axis.
The $p$-wave single-channel threshold wave function with mixed
$m$-character can thus be written as 
\begin{equation} 
\label{eq:fct-non-normee}
\phi(x,\theta)=\cos(\alpha)u_0(x) \cos(\theta) +\sin(\alpha)u_1(x) \sin(\theta)\,,
\end{equation}
where $u_0(x)$ and $u_1(x)$ are threshold radial components of the eigenfunctions of the two Hamiltonians with $\ell$=1. They have  the same asymptotic form $u_{|m|}(x)\equiv x^2 + \ldots$, cf. Paper I~\cite{paperI}. 
The angle $\alpha$ denotes the orientation of the dipole moments relative to the interparticle axis. 
More specifically, $\alpha$ determines the main orientation of the interparticle axis at large distance, where $u_0(x)$ and $u_1(x)$ are taken to be identical and equal to $x^2$. 
%
%%%%%%%%%%%%%%%%%%%%%%%%%%%%% figure 8 eta vs x%%%%%%%%%%%%%%%%%%%%%%%%%%%%%%%
\begin{figure}[tb]
  \centering
  \includegraphics[width=0.99\linewidth]{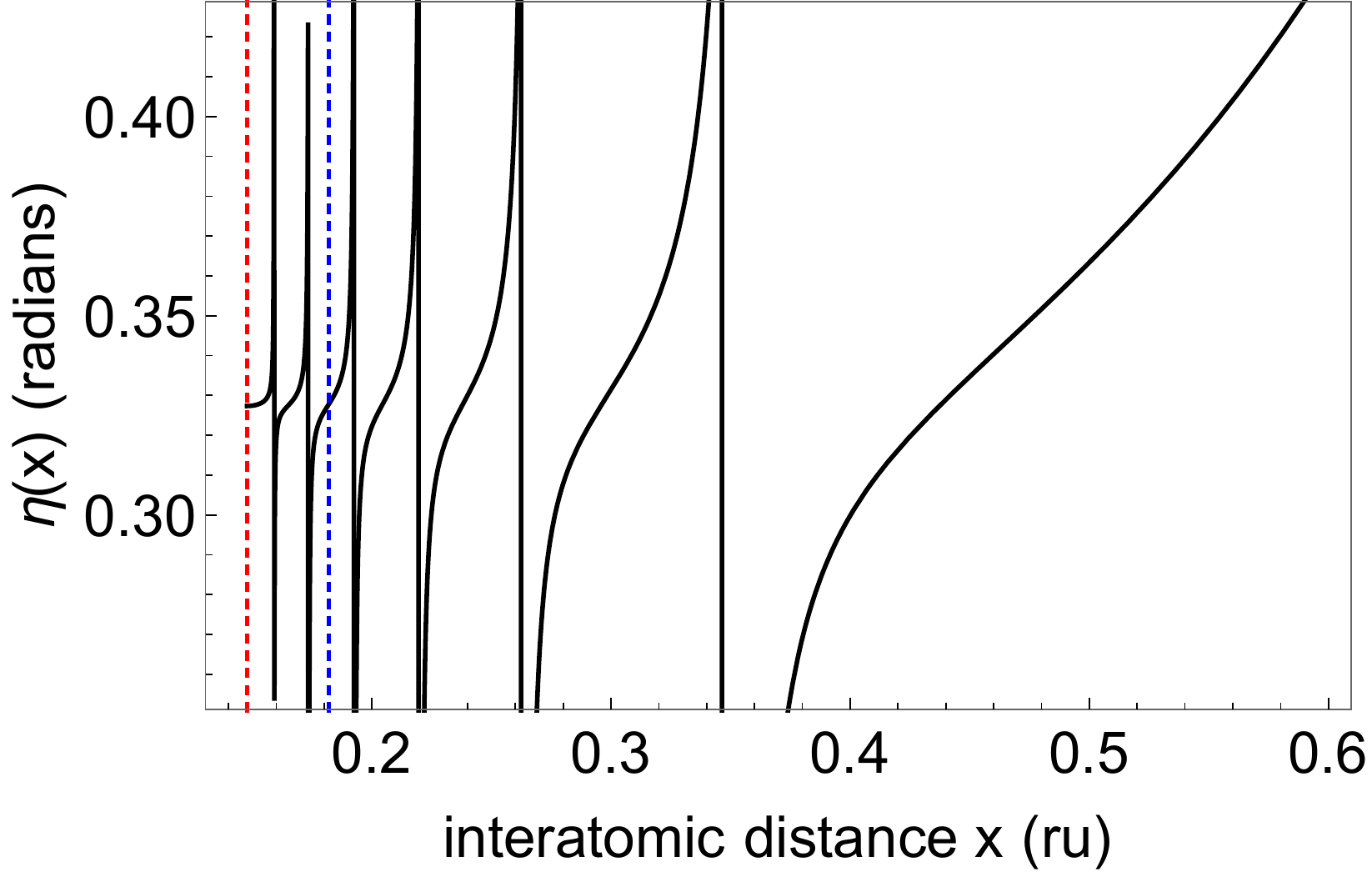}
  \caption{Dependence of the angle $\eta(x)$, characterizing the main orientation of the interparticle axis with respect to the field direction, on 
interparticle distance $x$ for ${\mathcal I}$=6~ru and
$x_{00}$=0.148~ru (which corresponds to an $s$-wave scattering length
of 0.651~ru). The vertical red dotted line indicates the position of
the repulsive wall $x_0(\mathcal E=0, \ell=1, \mathcal I)$ (see text)
and the blue one the position of the value of $x$ chosen in Fig.~\ref{fig:orientation}. 
The asymptotic value of $\eta (x)$ in this example is $\pi/4$. 
Calculations are performed in the single-channel $p$-wave model.
}
  \label{fig:orientation-wf}
\end{figure}
%%%%%%%%%%%%%%%%%%%%%%%%%%%%%%%%%%%%%%%%%%%%%%%%%%%%%%%%%%%%%%%%%%%%%%%%%%%%%%%
%
One can separate the function $\phi(x,\theta)$ into a radial part which depends only on the interparticle distance $x$ and is asymptotically equal to $x^2$ and an angular part,
\begin{subequations}
\label{eq:fct-normee}
\begin{eqnarray}
\nonumber 
\phi(x,\theta)&=&(\cos(\alpha)^2 u_0(x)^2+ \sin(\alpha)^2 u_1(x)^2)^{1/2} \\ \nonumber 
 & & 
\times\frac{\cos(\alpha)u_0(x) \cos(\theta) +\sin(\alpha)u_1(x) \sin(\theta)}{(\cos(\alpha)^2 u_0(x)^2+ \sin(\alpha)^2 u_1(x)^2)^{1/2}}\\\nonumber
\label{eq:fct-normee-1}
\\ \nonumber
&=&(\cos(\alpha)^2 u_0(x)^2+ \sin(\alpha)^2 u_1(x)^2)^{1/2} \\ 
 & &\times \cos(\theta-\eta(x)) 
\label{eq:fct-normee-2}
\end{eqnarray}
where
\begin{equation}
\tan(\eta(x))=u_1(x)/u_0(x)\times \tan(\alpha)\,.
\label{eq:angle-orientation}
\end{equation}
\end{subequations}
For a fixed interparticle distance $x$, the angle $\eta(x)$ is the angle for which the wave function presents
the maximum probability, i.e., it corresponds to the main orientation
of the interparticle axis.
In the asymptotic domain where $u_1(x)/u_0(x)$ is almost constant, $\eta(x)$ varies slowly and converges regularly toward its limit $\alpha$. 
In contrast, at short distance {($x< 0.5\,$ru ) where the attractive $-1/x^6$ potential dominates and the radial functions $u_{m=0,\pm 1}(x)$ are highly oscillatory, $\eta(x)$ changes rapidly. This is illustrated in Fig.~\ref{fig:orientation-wf} which displays $\eta$ as a function of interparticle distance $x$. As $x$ is decreased and approaches the nodal line, $\eta(x)$ approaches a value that depends on $\alpha$ and $x_{00}$,  apart from sudden variations at the nodes of the wave function.  This value amounts to $\sim0.33$~radians in the example of Fig.~\ref{fig:orientation-wf}.

%
%%%%%%%%%%%%%%%%%%%%%%%%%%%%% fig 9 eta vs x00 %%%%%%%%%%%%%%%%%%%%%%%%%%%%%%%%
%
\begin{figure}[tb]
  \centering
  \includegraphics[width=0.99\linewidth]{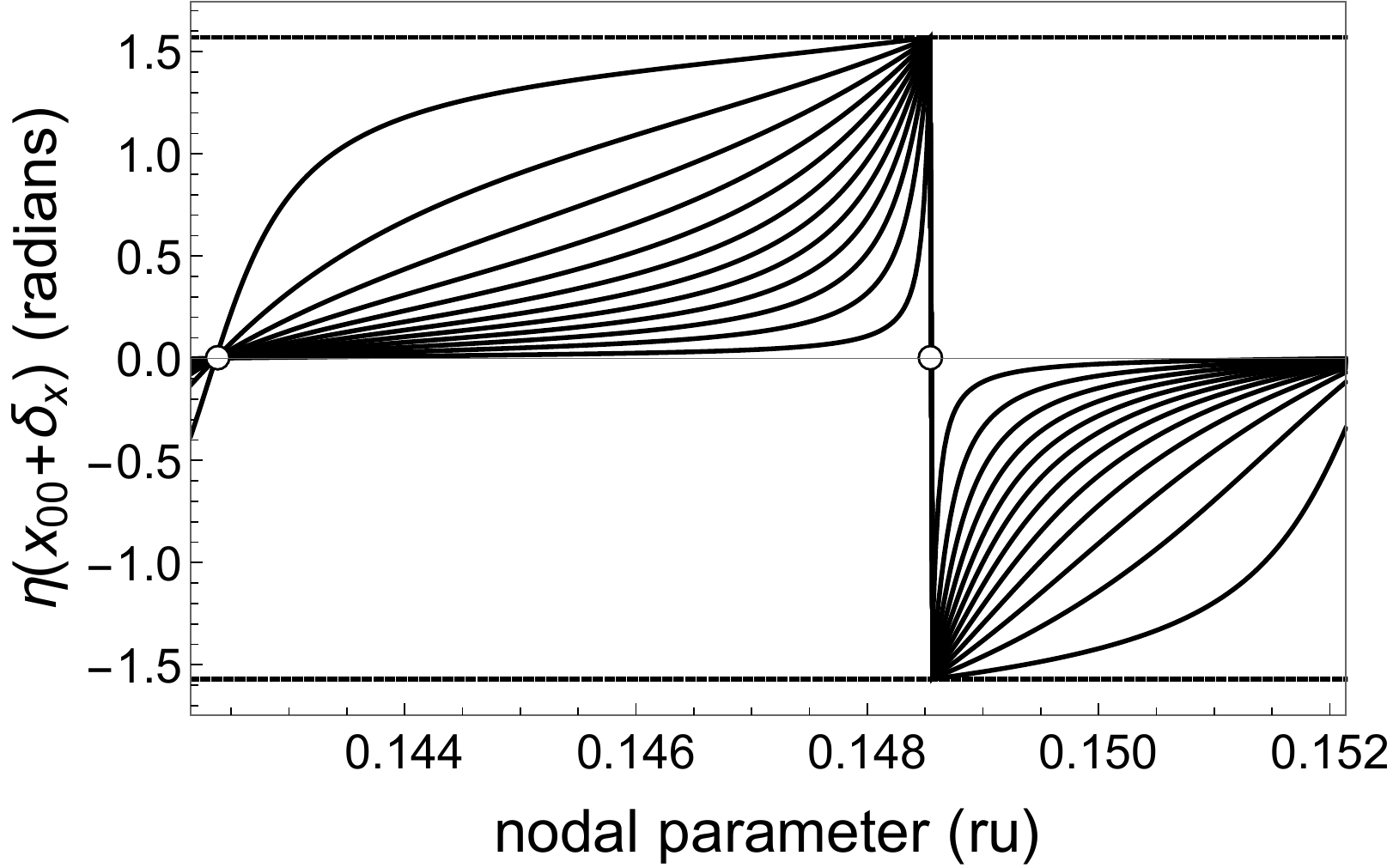}
  \caption{Angle $\eta(x_{00}+\delta_x)$, characterizing the main orientation 
of the interparticle axis with respect to the field direction at short
distance, as a function of the nodal parameter $x_{00}$ for various
values of the angle $\eta(x_{max})$=$\alpha$ at large distance
($x_{max}$=200~ru), with $\alpha$ varying from $\pi/24$ to
$11\pi/24$. The intensity is ${\mathcal I}$=6~ru and the value of
$\delta_x$,  $\delta_x=0.034\,$ru, is chosen such that
$x_{00}+\delta_x$ is small and does not coincide with a node of the
wave function. 
For a nodal parameter corresponding to a  bound state at threshold
with either $m$=0 or $|m|$=1, $\eta(x_{00}+\delta_x)$ is independent on the asymptotic orientation $\alpha$ and is equal to either 0 or $\pm\pi/2$.
The two small open circles indicate the corresponding positions of the
divergences of the generalized scattering volume for $m$=0 (left) and
$|m|$=1 (right). Calculations are performed in the single-channel
$p$-wave model.
}
  \label{fig:orientation}
\end{figure}
%%%%%%%%%%%%%%%%%%%%%%%%%%%%%%%%%%%%%%%%%%%%%%%%%%%%%%%%%%%%%%%%%%%%%%%%%%%%%%%%
%
The short-range behavior of $\eta(x)$ is further analyzed in
Fig.~\ref{fig:orientation} which shows the dependence  of
$\eta(x_{00}+\delta_x)$ on the nodal parameter $x_{00}$  for various
values of the asymptotic angle $\eta(x_{max})$=$\alpha$ (evaluated
here at $x_{max}=200\,$ru), with $\alpha$ varying from $\pi/24$ to
$11\pi/24$. 
The value of $\delta_x$ is chosen such that $x_{00}+\delta_x$ is small
and does not coincide with a node of the $u_{0,\pm1}(x)$  wave functions. The two constant cases correspond to $\alpha$=0 and $\alpha=\pi/2$.
The short-range dependence  of $\eta (x)$ on $x_{00}$ was obtained by
calculating the slopes $D_m(x_{0})=u'_m(x_{0})$ of the two solutions
at the position of their energy-, intensity- and $\ell$-dependent node
$x_{0}(\mathcal E, \ell, \mathcal I)$ with $\ell=1$ and $\mathcal E=0$. This is explained in App.~\ref{sec:eta}.
A remarkable observation can be made in the case when the scattering
volume diverges, with the corresponding
values of $x_{00}$ indicated by the open circles in
Fig.~\ref{fig:orientation}. Then $\eta(x_{00}+\delta_x)$ takes the
same value, independent of its asymptotic value $\alpha$, as is
evident from all curves in  Fig.~\ref{fig:orientation} coinciding.
This is in contrast to the case when the scattering volume remains finite, in which case the short-range value of $\eta$ does depend on the asymptotic value.
Next, we will examine whether the special behavior of the orientation
of the interparticle axis relative to the dipole moments in the case
of a diverging scattering volume still appears when the coupling between different partial waves is properly accounted for.

%
%ooooooooooooooooooooooooooooooooooooooooooooooooooooooooooooooooooooooooooooooo
\subsection{Orientation of the interparticle axis at short internuclear distance: several channels}
\label{subsec:orient-multi}
%ooooooooooooooooooooooooooooooooooooooooooooooooooooooooooooooooooooooooooooooo
%
%
%%%%%%%%%%%%%%%%%%%%%%%%%%%%% figure 11 distribution piquee %%%%%%%%%%%%%%%%%%%%
\begin{figure}[tb]
  \centering
  \includegraphics[width=0.6\linewidth]{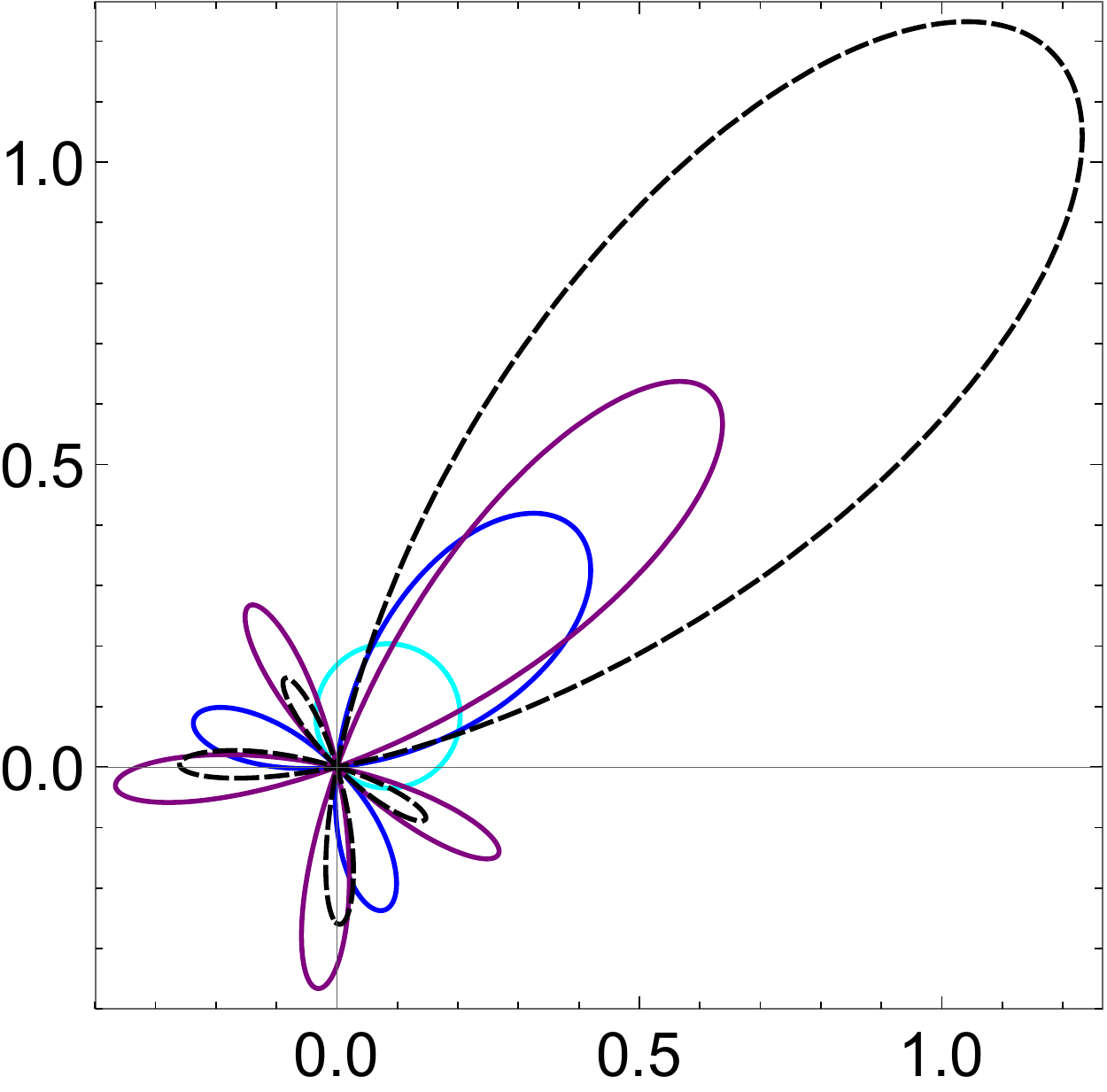}
  \caption{Polar plot of the asymptotic wave function~\eqref{eq:phi-infini}, for three channels with odd $\ell$ values, $\theta_0=\pi/4$ and $\phi=0$. Dashed black line: total wave function. Cyan, blue and purple lines: partial wave functions with $\ell$=1, 3 and 5.}
  \label{fig:orientation-x-infini}
\end{figure}
%%%%%%%%%%%%%%%%%%%%%%%%%%%%%%%%%%%%%%%%%%%%%%%%%%%%%%%%%%%%%%%%%%%%%%%%%%%%%%%%
%
To analyze the role of the scattering volume for the orientation of
the interparticle axis in a multi-channel treatment, we consider a
given asymptotic orientation, and look at the angular behavior of the
corresponding wave function as $x$ decreases. This approach is
motivated by the fact that any actual situation can be described by a
superposition of wave functions with given asymptotic behavior. We start from the following general expression of the Dirac $\delta$-function:
\begin{equation} 
\label{eq:delta-teta}
\frac{\delta(\theta-\theta_0)}{ \sin(\theta_0)}=\sum_{\ell=0}^{\infty} \sum_{m=-\ell}^{\ell} Y_{\ell,m}(\theta,\phi) Y_{\ell,m}^\star(\theta_0,\phi)\,.
\end{equation}
At large distance, the wave function providing the best representation of a given orientation $\theta_0$ of the interparticle axis with respect to the laboratory fixed $Z$ axis can be written as
\begin{equation} 
\label{eq:phi-infini}
f_{asym}(\theta,\phi)=\sum_{\ell=\ell_{min}}^{\ell_{max}} \sum_{m=-\ell}^{\ell} Y_{\ell,m}(\theta,\phi) Y_{\ell,m}^\star(\theta_0,\phi)\,. 
\end{equation}
We limit the sum over $\ell$ to odd values and restrict $\phi$ to zero due to symmetry, as in the previous subsection. Figure~\ref{fig:orientation-x-infini} shows the asymptotic wave function for the example of $\theta_0=\pi/4$, obtained by including three values  of $\ell$ and all corresponding $m$ values in the calculation. As expected, the wave function points towards $\pi/4$ and, as also expected, higher $\ell$-waves are required to properly describe the orientation. 

To study the $x$-dependence  of the angular behavior of the wave function, we solve the Schr\"odinger equation for the three values of $\ell$ and all corresponding $m$ values. We use the same method of inward integration as described in Paper I~\cite{paperI} and obtain a set of radial wave functions $u_{\ell,\ell'}^m(x)$, where $\ell$ denotes the channel associated to the physical solution and $\ell'$  refers to the channel in which the integration starts. At large distance, taken to be $x_{max}$, the interaction between the different channels is small, and $u_{\ell,\ell'}^m(x_{max})$ is roughly proportional to a Kronecker $\delta_{\ell,\ell'}$. The complete wave function associated to given asymptotic conditions is thus given by
\begin{eqnarray}
\label{eq:wf}
f(x,\theta,\phi)=\sum_{\ell=\ell_{min}}^{\ell_{max}} \sum_{\ell'=\ell_{min}}^{\ell_{max}} \sum_{m=-\ell}^{\ell} && u_{\ell,\ell'}^m(x) \\\nonumber
&&\; Y_{\ell,m}(\theta,\phi) Y_{\ell',m}^\star(\theta_0,\phi) \,,
\end{eqnarray}
where $\ell$ only takes odd values.
We calculate separately the different $\ell$-components of the wave function, 
\begin{equation} 
\label{eq:wf-ell}
f_{\ell}(x,\theta,\phi)=\sum_{\ell'=\ell_{min}}^{\ell_{max}}\sum_{m=-\ell}^{\ell} Y_{\ell,m}(\theta,\phi) Y_{\ell',m}^\star(\theta_0,\phi)u_{\ell,\ell'}^m(x) \,,
\end{equation}
and their norms,
\begin{equation} 
\label{eq:norme}
N_{\ell}(x,\phi)^2= \sum_{\ell'=\ell_{min}}^{\ell_{max}}\sum_{m=-\ell}^{\ell} |Y_{\ell',m}(\theta_0,\phi)u_{\ell,\ell'}^m(x)|^2\,.
\end{equation}
%
%
%%%%%%%%%%%%%%%%%%%%%%%%%%%%% figure 12 hors divergences %%%%%%%%%%%%%%%%%%%%%%% 
\begin{figure}[tb]
  \centering
  \includegraphics[width=0.99\linewidth]{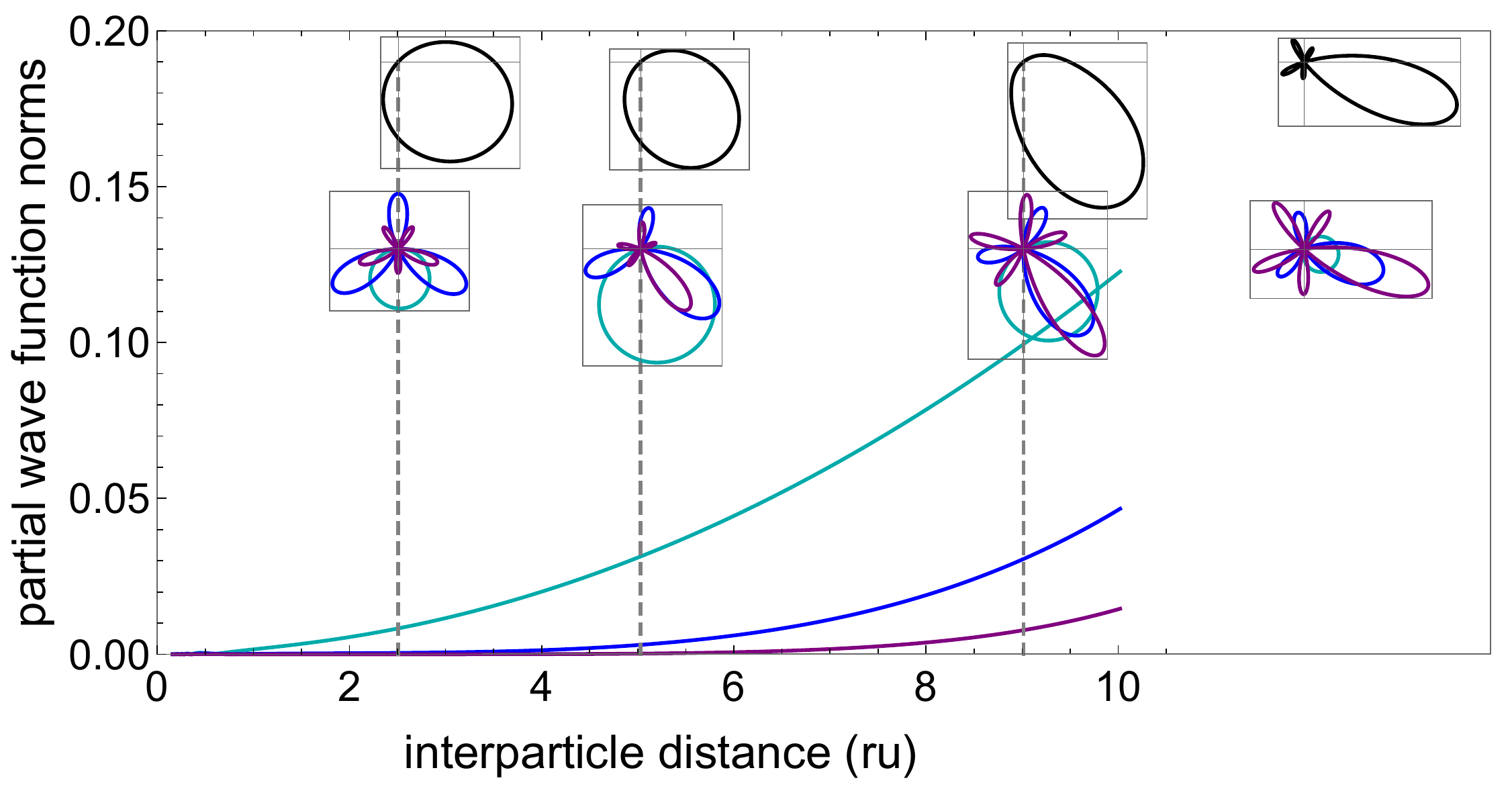}
  \caption{Partial wave norms (cf. Eq.~\eqref{eq:norme}, 
cyan, blue, and purple curves for $\ell=$1, 3, 5)
 as a function of interparticle distance, for $\theta_0=-0.3\pi$ and a nodal parameter that corresponds to a generalized scattering volume far from the
$\widetilde{\ell}$=1 poles ($x_{00}$=0.147~ru or, resp. $a$=0.494623~ru).
%%% el3: specify \mathcal I
The insets show the angular behavior of the scattering wave function
for several $x$ whose positions are indicated by the vertical gray
dashed lines. The angular behavior of the total wave function,
Eq.~\eqref{eq:wf}, is depicted in black, at the top. The polar plots
corresponding to the  partial wave functions,
Eq.~\eqref{eq:wf-ell},  are also shown, at the bottom with the
asymptotic one to the right.
For $x$ smaller than about 2.5~ru, the angle for which the probability
is maximum becomes roughly fixed, with a value depending on the
asymptotic orientation (here about -$\pi/4$).  The calculation was performed with three values of $\ell$ and all corresponding $m$ values.}
  \label{fig:orientation-hors-div}
%%% chr: the insets should not overlap! also, it would be better not
%%% to have an inset overlap with any of the curves in the main
%%% graph. 
%% ac: I will do it maybe later
\end{figure}
%%%%%%%%%%%%%%%%%%%%%%%%%%%%%%%%%%%%%%%%%%%%%%%%%%%%%%%%%%%%%%%%%%%%%%%%%%%%%%%%%
%
In general, when the absolute value of the generalized scattering volume is not too large, the partial wave functions are elongated according to the expected orientation at large distances and the evolution of the orientation with decreasing $x$ does not present a spectacular behavior. This is illustrated in Fig.~\ref{fig:orientation-hors-div}. The only notable point is that the orientation becomes fixed at short distance, with a direction that depends on the asymptotic orientation, just as in the single channel case. 

%
%%%%%%%%%%%%%%%%%%%%%%%%%%%%% figure 13 divergences l=1  %%%%%%%%%%%%%%%%%%%%%%%%
\begin{figure}[tb]
  \centering
  \includegraphics[width=0.99\linewidth]{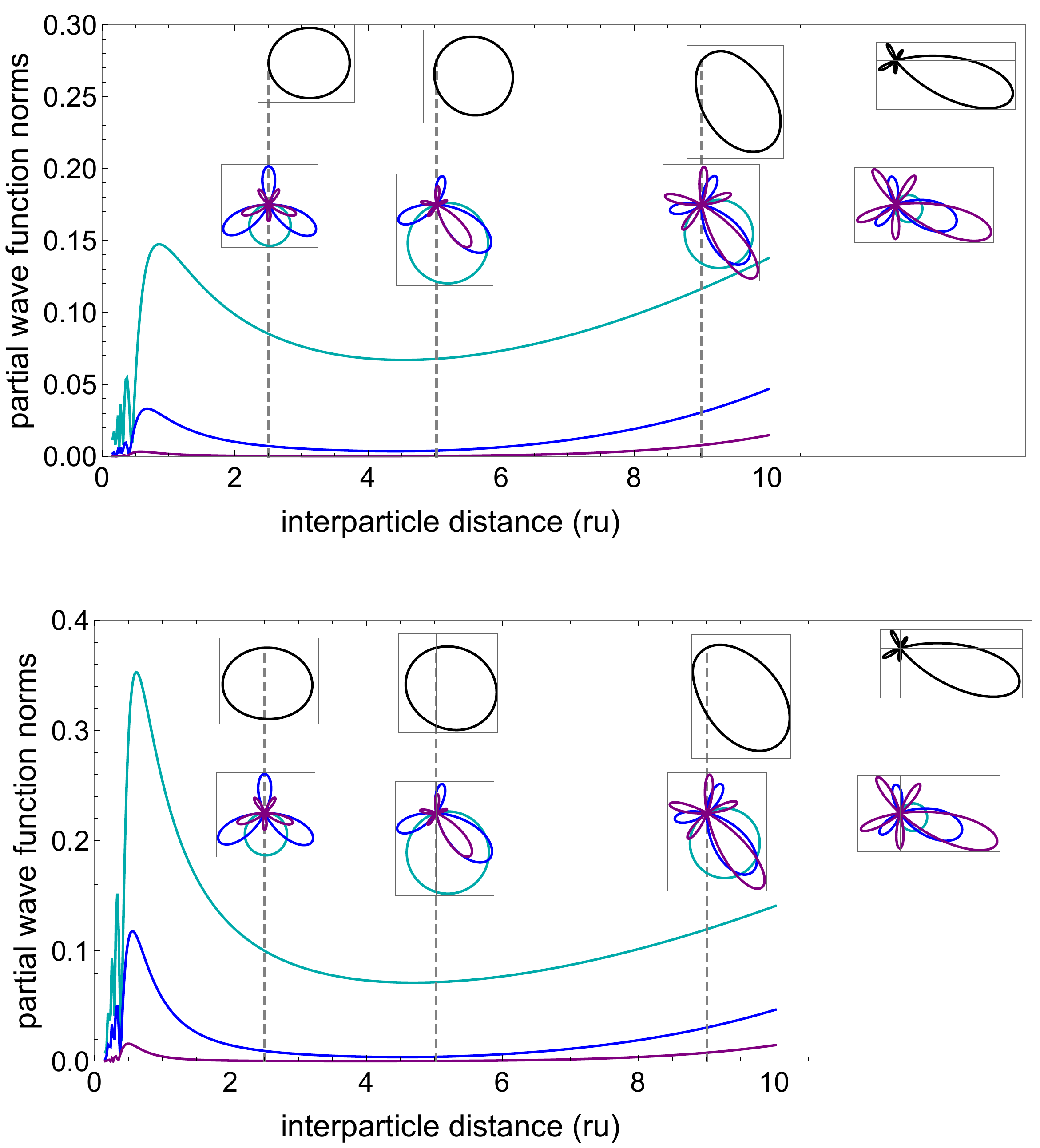}
  \caption{Same as Fig.~\ref{fig:orientation-hors-div}, but for a
    nodal parameter corresponding to a generalized scattering volume
    close to a pole with $\widetilde{\ell}$=1. Top: pole with $m$=0,
    $x_{00}$=0.142906~ru (s-wave scattering length $a$=-1.31436~ru),
    bottom: pole with $|m|$=1, $x_{00}$=0.149140~ru
    ($a$=0.87585~ru). For $x$ smaller than about 2.5~ru, the angle for
    which the probability is maximum becomes more or less fixed, with
    the direction not depending on the asymptotic orientation (0 for
    the divergence with $m=0$ and $-\pi/2$ for that with $|m|=1$). 
  }
  \label{fig:orientation-resos-1}
\end{figure}
%%%%%%%%%%%%%%%%%%%%%%%%%%%%%%%%%%%%%%%%%%%%%%%%%%%%%%%%%%%%%%%%%%%%%%%%%%%%%%%%%
The situation is quite different when the scattering volume is close to one of its poles (for given $\ell,m$), cf. Fig.~\ref{fig:orientation-resos-1} for two poles with $\ell$=1.
In this case, the partial wave norms, especially the one corresponding to the $\ell,m$ value of the pole, have a large maximum a short distance. Moreover, the orientation of the interparticle axis takes a fixed direction at short range, 0 or $\pi$ for $m$=0 and $\pm \pi/2$ for $|m|$=1, depending on  the asymptotic orientation. 
In the first case, the dipoles are head-to-tail whereas in the latter one, the dipoles become roughly perpendicular to the interparticle axis.

%%%%%%%%%%%%%%%%%%%%%%%%%%%%% figure 14 divergence l=5 m=1  %%%%%%%%%%%%%%%%%%% 
\begin{figure}[tb]
  \centering
  \includegraphics[width=0.99\linewidth]{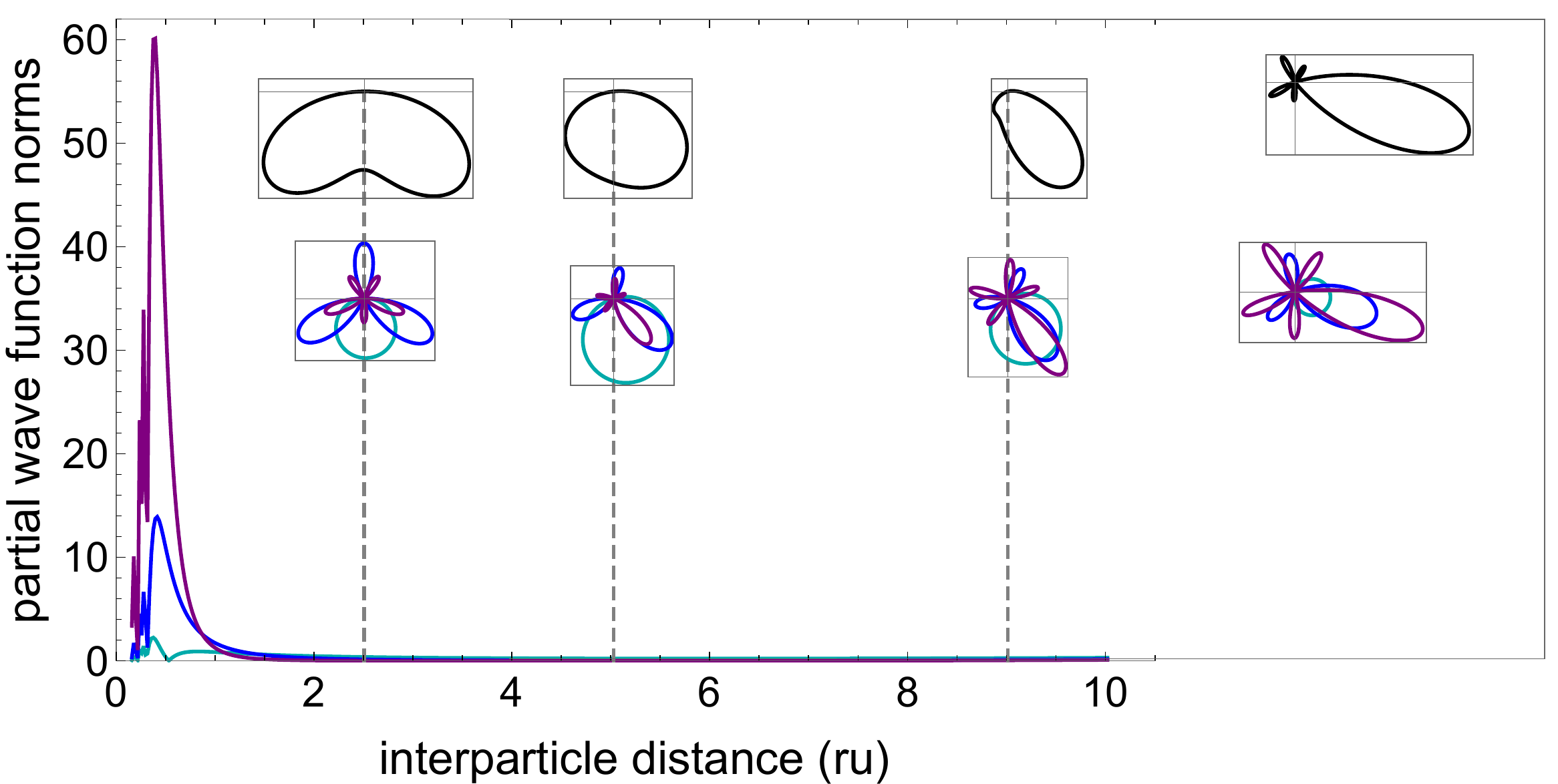}
  \caption{Same as Fig.~\ref{fig:orientation-hors-div}, but for a
    nodal parameter corresponding to a generalized scattering volume
    close to a pole with  $\widetilde\ell$=5, $|m|$=1
    ($x_{00}$=0.150116~ru or, resp., $a$=1.19953~ru). For $x$ smaller
    than about 2.5~ru, the average angle $\theta$ becomes more or less
    fixed,  at a value close to $\pm \pi/2$, not depending on the asymptotic orientation.}
  \label{fig:orientation-reso-5}
\end{figure}
%%%%%%%%%%%%%%%%%%%%%%%%%%%%%%%%%%%%%%%%%%%%%%%%%%%%%%%%%%%%%%%%%%%%%%%%%%%%%%%%
%
A similar behavior is observed for poles with other values of $\ell$
but the resonance character of the wave function may become even more
pronounced. This is illustrated in Fig.~\ref{fig:orientation-reso-5}
which shows the example of a pole with $\ell=5$, $|m|=1$. The main
direction at short distance is $\pi/2$, as for the pole $\ell=1$,
$|m|=1$, i.e., the dipoles are perpendicular to the interparticle
axis. However, the relative importance of the partial waves is quite
different and the maximum of the partial wave norms with a larger
$\ell$ occurs at shorter distance, since the rotational  barrier for
$\ell=5$ is  higher and located at a smaller separation than the $\ell=1$ barrier.

In conclusion, close to a singularity of the generalized scattering volume, the main orientation at short interparticle distance is fixed, irrespective of the specific experimental conditions (except for the pancake or needle-shaped samples). 
So controlling the generalized scattering volume, either by tuning non-resonant light or by choosing an effective dipole length for aligned permanent dipoles, does not only affect the interaction strength of the scattering partners but also their orientation. While this is expected for collisions of polar molecules, it is less obvious for scattering in the presence of non-resonant light.

%
%ooooooooooooooooooooooooooooooooooooooooooooooooooooooooooooooooooooooooooooooo
\subsection{Fixed orientation of the internuclear axis}
\label{subsec:orient-fixed}
%ooooooooooooooooooooooooooooooooooooooooooooooooooooooooooooooooooooooooooooooo
%
%
%%%%%%%%%%%%%%%%%%%%%%%%%% figure 15 melange %%%%%%%%%%%%%%%%%%%%%%%%%%%%%%%%%%%
\begin{figure*}[tb]
  \centering
  \includegraphics[width=0.99\linewidth]{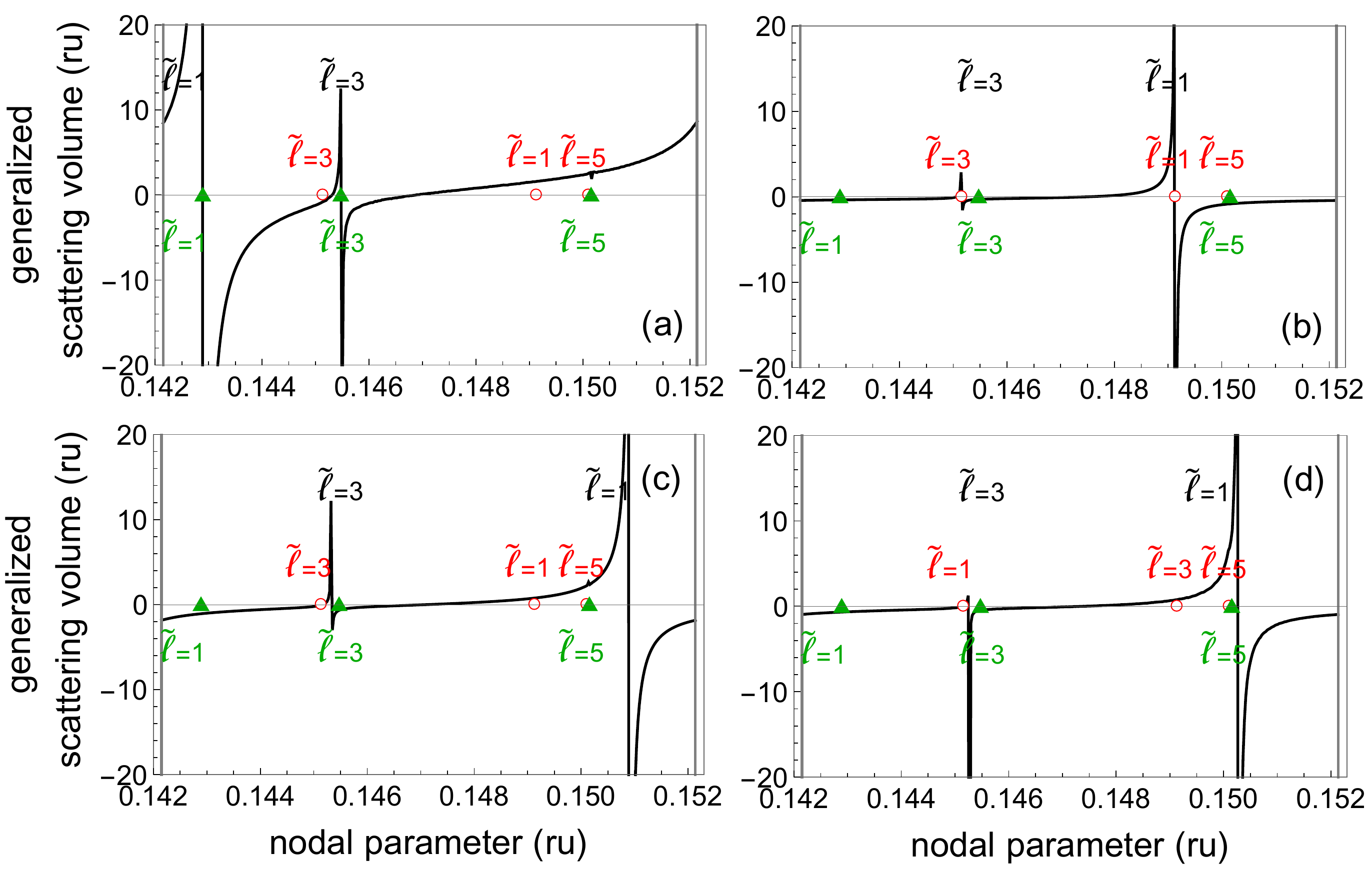}
  \caption{\label{fig:fixed-orientation} Dependence of the generalized field-dressed scattering volume on the nodal parameter $x_{00}$ for a fixed  orientation of the interparticle axis, with a fixed mixing of $m$=0 and $m$=$\pm 1$ states characterized by the values of $\cos^2\alpha$ and $\sin^2\alpha$ in the Hamiltonian~\eqref{eq:H-alpha}. (a): $\alpha$=0, only $m$=0; (b) $\alpha$=$\pi/2$, only $m$=$\pm 1$; (c): $\alpha$=$\pi/4$, equal mixing of the two values; (d): $\alpha$=$\alpha_{QL}$, with $(cos^2(\alpha),sin^2(\alpha))$=$(1/3,2/3)$. The red open circles (green triangles) indicate the divergence of the generalized scattering
volume for pure $m$=0 ($|m|$=1) states. The vertical gray lines indicate the $x_{00}$ values corresponding to infinite field-free $s$-wave scattering length. The calculations are performed for three channels and $\mathcal I$=6~ru. 
	}
\end{figure*}
%%%%%%%%%%%%%%%%%%%%%%%%%%%%%%%%%%%%%%%%%%%%%%%%%%%%%%%%%%%%%%%%%%%%%%%%%%%%%%%%%
We will now consider the situation where the direction of the
interparticle axis is fixed by geometrical constraints due to the trap, such as those encountered in a disk or needle sample.
For a given orientation $\alpha$ of the interparticle axis with respect to the common dipole direction, the Hamiltonian can be written as
\begin{equation}
H_{\alpha}=\cos^2(\alpha)\,\, H_0 + \sin^2(\alpha)\,\, H_1 \, ,
\label{eq:H-alpha}
\end{equation}
where the  $m$-dependent single-channel Hamiltonians $H_m$ are obtained from the 2D Hamiltonian $H$ in Eq.~\eqref{eq:2D_Hamil} as
\begin{equation}
H_m(x,\ell,\ell')=\langle \,Y_\ell^m | H |Y_{\ell'}^m \,\rangle\,.
\label{eq:H-ell}
\end{equation}
In a calculation accounting for $n$ partial $\ell$-waves, $H_{\alpha}$ becomes a $n\times n$ matrix.
The effective potential $V_{eff}$ that governs the generalized scattering volume is the diagonal matrix element for the $p$-wave channel, equal to
\begin{equation}
V_{eff}(x)=\frac{2}{x^2} - \frac{1}{x^6} - \mathcal I \frac{4 \cos^2(\alpha) - 2\sin^2(\alpha)}{15x^3}\,.
\label{eq:H-alpha-p}
\end{equation}
We analyze the behavior of the generalized field-dressed scattering volume for four different values of $\alpha$ in Fig.~\ref{fig:fixed-orientation}. 
These values correspond to the case of $m$=0 and $m$=1 states alone in
Fig.~\ref{fig:fixed-orientation}(a) and (b), respectively, an equal mixture of $m$=0 and $|m|$=1 states in Fig.~\ref{fig:fixed-orientation}(c), and to the case (d), 
where $\alpha$=$\alpha_{QL}$ and the potential becomes quasi-long range (QL)~\cite{MarinescuPRL98}}. In this latter case, $\cos^2(\alpha)=1/3$, $\sin^2(\alpha)=2/3$. This means that the $1/x^3$ term due to the non-resonant light (or dipole-dipole interaction) disappears from the diagonal term of the Hamiltonian. The quasi-long range character of the interaction obtained in this case is analogous to that in the problem of non-resonant light control of the $s$-wave scattering length for even parity states~\cite{CrubellierPRA17}.

Each panel in Fig.~\ref{fig:fixed-orientation}
displays essentially two divergences of the scattering volume, which
can be labeled by $\widetilde{\ell}$=1 and $\widetilde{\ell}$=3. While
this is expected in the case (a) and (b) where there is no $m$-mixing,
it is more surprising in the other cases. These divergences can in all cases
be labeled by $\widetilde{\ell}=1$ and $\widetilde{\ell}=3$ (the $\widetilde\ell$=5 divergences, too narrow, are not visible here). 
Note that the positions of the $\widetilde{\ell}=1$ divergences vary notably with $\alpha$:
The $\tilde \ell=1$ resonances of the two pure cases (top part of
Fig.~\ref{fig:fixed-orientation}) are located at very distant $x_{00}$
values so that the $\tilde \ell=1$ pole for mixed $m$  (in
Fig.~\ref{fig:fixed-orientation}(c)) lies between the  $m=0$ pole and
the  $|m|=1$ pole of the next interval of $x_{00}$ values (remember the quasi-periodicity  of the nodal line model with $x_{00}$). 
For $\widetilde \ell$=3, the two 'pure' poles are very close one to
the other and are not appreciably displaced also in the equal mixing case. 
Suprisingly, the quasi-long range case in Fig.~\ref{fig:fixed-orientation}(d)
does not differ in any essential way from the other three cases.

\begin{figure}[tb]
  \centering
  \includegraphics[width=0.99\linewidth]{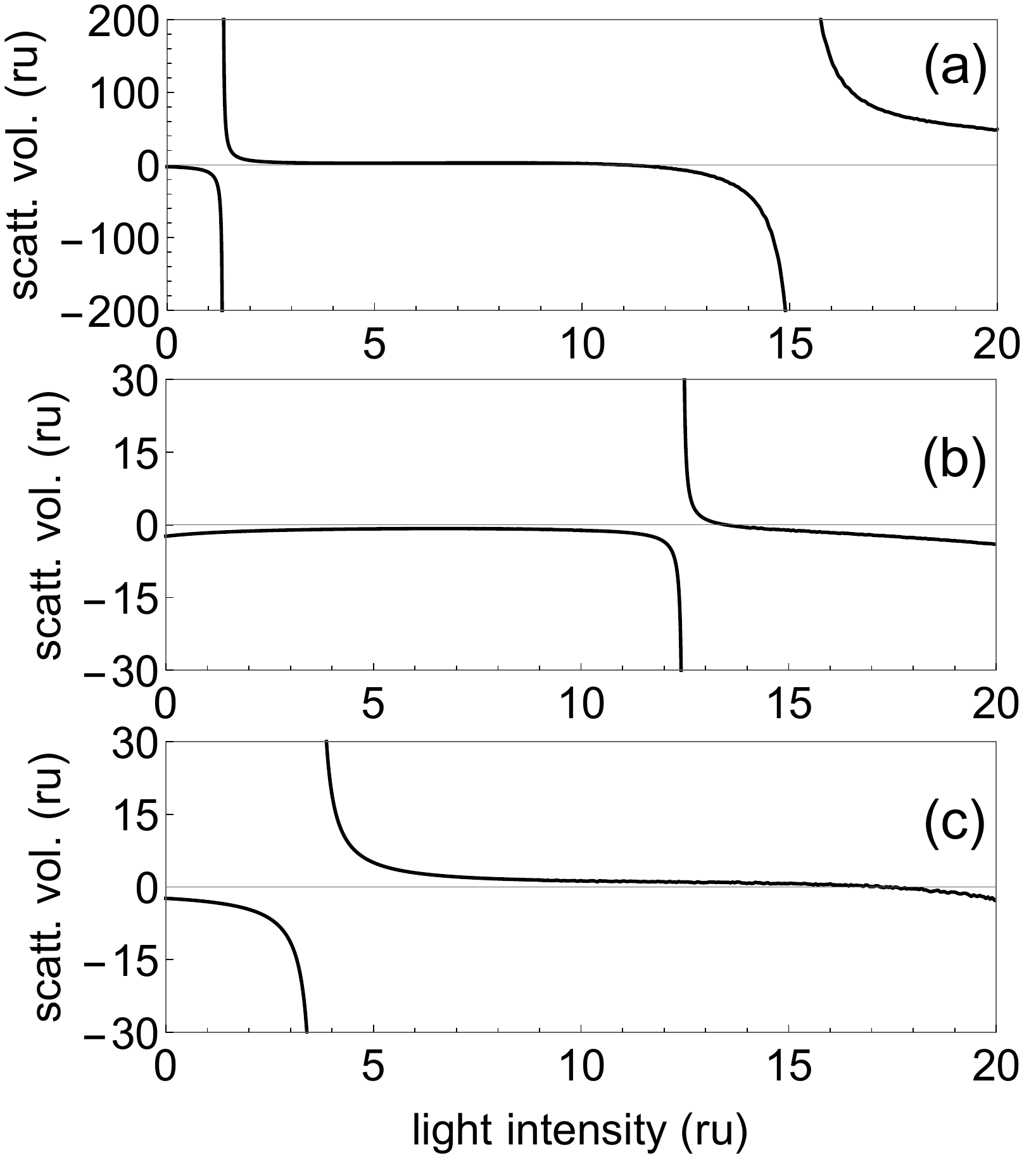}
  \caption{\label{fig:orientation-vs-I} 
    Dependence of the generalized field-dressed scattering
    volume on the light intensity for a given orientation of the
    interparticle axis.
    (a): $\alpha=0$, i.e., $m=0$ alone; (b) $\alpha=\pi/2$, i.e., 
    $m=\pm 1$; (c): $\alpha=\pi/4$, equal mixing of $m=0$ and $m=\pm 1$ states. The
    calculations are performed for a field-free s-wave scattering length
    equal to 1.16$\,$ru, within a three channel model.
%%% chr: does this scattering length correspond to a real example? if so, which?
%%% yes triplet 41K87Rb dans la ref 19. I admit that this example is not very interesting...
  }
\end{figure}
%%%%%%%%%%%%%%%%%%%%%%%%%%%%%%%%%%%%%%%%%%%%%%%%%%%%%%%%%%%%%%%%%%%%%%%%%%%%%%%%
Figure~\ref{fig:orientation-vs-I} illustrates the dependence of the generalized scattering volume on the non-resonant light intensity for three different asymptotic orientations $\alpha$. For $\alpha=0$, shown in  Fig.~\ref{fig:orientation-vs-I}(a), only the $m=0$ term of the Hamiltonian~\eqref{eq:H-alpha} contributes, and a singularity is observed at $\mathcal I=1.35\,$ru with a width of 3.68$\,$ru. The second singularity in 
Fig.~\ref{fig:orientation-vs-I}(a) is a consequence of the
quasi-periodicity of the model. For $\alpha=\pi/2$, only the $|m|=1$
term in Eq.~\eqref{eq:H-alpha} comes into play, and the singularity in
Fig.~\ref{fig:orientation-vs-I}(b) is found at much higher intensity,
$\mathcal I=12.45\,$ru (with a width of 1.21$\,$ru). Whereas the
position of the pole in the equal mixing case ($\alpha=\pi/4$), cf.  Fig.~\ref{fig:orientation-vs-I}(c), is intermediate
between the positions in the two pure-$|m|$ cases, at $\mathcal I=3.64\,$ru, the dependence of the width (equal to 7.08$\,$ru in the mixed case)
on orientation is less obvious to explain. This is due to the strong dependence of the width on the intensity which is increasing for $m=0$ and decreasing for $|m|=1$ in the pure-$|m|$ cases. All of the singularities shown in Fig.~\ref{fig:orientation-vs-I} are characterized by $\widetilde\ell=1$.
The poles for $\widetilde\ell=3$ appear for smaller $s$-wave scattering
lengths, cf. Fig.~\ref{fig:scatt-any}, and those for $\widetilde\ell=5$ are too narrow to have been resolved in the present calculation.

%
%-------------------------------------------------------------------------------
\section{Conclusions}
\label{sec:conclusion}
%-------------------------------------------------------------------------------
We have studied non-resonant light control of the $p$-wave scattering
volume characterizing collisions of identical spin-polarized fermions
at very low energy. To this end, we have employed an asymptotic model~\cite{LondonoPRA10,CrubellierPRA17} to describe the
low-energy collisions. This is justified by the predominance of
long-range forces at these energies. The short-range interactions are
represented by a single parameter, the nodal parameter, in the
asymptotic model. It can be fixed if the field-free $s$-wave
scattering length of the collision partners is known~\cite{LondonoPRA10}. 
Since the interaction with the non-resonant light scales
asymptotically as $1/R^3$ with the interparticle distance, it was
necessary to first generalize the definition of the scattering volume, cf. Paper I~\cite{paperI}. 

For free particles or weak confinement, we have determined when
singularities of the field-dressed generalized $p$-wave scattering volume occur as a
function of the non-resonant light intensity and the field-free
$s$-wave scattering length, resp. the nodal parameter, i.e. the specific
colliding pair. The singularities indicate the appearance of a bound
state at threshold and correspond to infinitely strong
interactions between the identical spin-polarized fermions. As a
result, for a given pair of particles, intensities close to a
singularity offer the most efficient control over the collisions. The necessary
intensities are of the order of 1$\,$GW/cm$^2$, with the lowest
intensities required for strongly polarizable particles with large
reduced mass. Such intensities are challenging but feasible with current
experimental technology. 

Our findings are quite similar to those of our earlier
study on using non-resonant light to control the $s$-wave scattering length for
identical bosons or non-polarized fermions~\cite{CrubellierPRA17}. The
main difference is that, at least for certain species, various efficient
means to control the $s$-wave scattering length exist, most notably
magnetic field control of Feshbach resonances~\cite{ChinRMP10}. In contrast,
external field control of the $p$-wave scattering volume has remained
an open goal. This may make the generation of the required
non-resonant light intensities a worthwhile experimental endeavor.

We have also considered non-resonant light control of $p$-wave
collisions for strongly confined particles, assuming an isotropic,
harmonic 3D trap. In this case, the asymptotic phase shift of the
scattering wave function is replaced by an energy shift of the trap
states. The energy shift for the odd-$\ell$ trap states can be directly related
to the scattering volume of free collisions. The same is true for
$s$-wave scattering where the even-$\ell$ trap state energy shift is
correspondingly related to the scattering length.

When the intensity of the non-resonant light is varied in a range
where we expect the field-dressed generalized scattering volume for
free collisions to diverge, the trap states get strongly
perturbed. The perturbation may be so strong as to permit 
up- or downward climbing of the trap ladder. Under these conditions,
it will also be possible to create bound molecular states by slowly
varying the non-resonant light intensity. In the vicinity of the divergences, 
the trap  state energy shifts can be  directly
related to the generalized scattering volume.  In contrast, away from the resonance, the trap states keep their character. 

Being of essentially $\ell=1$ character (even in the presence of
non-resonant light), $p$-wave scattering implies a mixing of the
states with $m$=0 and $|m|$=1. The relative weights of the $m$-states fix
the most probable relative orientation of light polarization and
interparticle axis. In a single channel approximation, the
orientation for two particles at close range tends to a more or less
fixed value. This value generally depends on the asymptotic
orientation, except in the proximity of a divergence of the
generalized scattering volume. In the latter case, the short-range
orientation is such that the particles are approximately head to tail
if the pole corresponds to an attractive interaction
($m$=0). Conversely, if the pole corresponds to a repulsive
interaction ($|m|$=1), the interparticle axis becomes approximately
perpendicular to the light polarization. Coupled channel calculations
with three values of $\ell$ and all corresponding $m$-values have confirmed and
amplified these results. While in an experiment the orientation of the 
dipole moments (induced or permanent) can be imposed by an external field, it is in general non-trivial to fix the orientation of the interparticle axis and thus the weights of the $m$-states which determine the anisotropic deformation of an expanding cloud~\cite{StuhlerPRL05}.

In the present calculations,  we have used 'universal' nodal lines
with a single energy-, partial wave- and intensity-dependent parameter, the nodal parameter, which in turn only  depends on the
field-free $s$-wave scattering length ~\cite{CrubellierNJP15a}.
Our predictions of the non-resonant light intensity required to
observe these phenomena could be made more precise by a better account
of the short-range interactions, using realistic nodal lines adjusted to experimental data.  This modification will be important in particular for collisions at somewhat higher energy, for example when studying shape resonances~\cite{LondonoPRA10,CrubellierNJP15a,CrubellierNJP15b}.

The asymptotic model used here to describe the interaction of
polarizable particles  with non-resonant light is not restricted to
this physical setting. Most importantly, collisions of aligned polar particles
at ultralow energies yield
the same asymptotic Hamiltonian. It is merely the meaning of the
reduced units that changes, and the anisotropic $1/R^3$-interaction is
due to the dipole moments of the colliding particles. As a consequence, the
calculations presented here also predict the $p$-wave scattering
volume (without any external field) as a function of the dipole moments. Of course, in this case,
the effective dipolar interaction strength cannot as easily be tuned
as in the case of non-resonant light control.

Given the generality of the asymptotic model with anisotropic
$1/R^3$-interaction, a natural extension of the present work would be
to explore the dynamics of two interacting ultracold dipoles confined
in an only axially symmetric harmonic potential. The investigation of
eigenenergies and eigenfunctions is possible for different geometries
of the trapping potential, from a pancake-shaped to a cigar-shaped
trap, all the way down to quasi-two-dimensional regimes.  The trap
geometry is known to influence the stability and excitations of
dipolar gases~\cite{YiPRA00,YiPRA02}. In particular, one could design
sample shapes that impose a specific orientation, or in other words,
fix the weights of the $m$-states. This is intriguing
in view of the different character of the $p$-wave scattering volume
singularities for $m$=0 and $|m|$=1 states that we have observed
here.  A further
extension would be to consider anharmonic traps.

%
%-------------------------------------------------------------------------------
\begin{acknowledgments}
Laboratoire Aim\'{e} Cotton is   
"Unit\'e mixte UMR 9188 du CNRS, de l'Universit\'e Paris-Sud, de 
l'Universit\'e Paris-Saclay et de l'ENS Cachan", member of the
"F\'{e}d\'{e}ration Lumi\`{e}re Mati\`{e}re" (LUMAT, FR2764) and of
the "Institut Francilien de Recherche sur les Atomes Froids" (IFRAF).
R.G.F. gratefully acknowledges financial support by the Spanish
Project No.  FIS2014-54497-P (MINECO), and by the Andalusian research group FQM-207. 
\end{acknowledgments}

\appendix

\section{Scaling parameters connecting trap energy shift and generalized scattering volume}
\label{sec:app}

We present here a general procedure to determine the parameters of the
linear transformation~\eqref{eq:osc-shift} connecting the trap energy
shift and field-dressed generalized scattering volume discussed in Sec.~\ref{subsec:traps-x00}.

Since $A$ corresponds to a constant shift of the harmonic oscillator levels due to the presence of the interactions, cf. Eq.~\eqref{eq:osc-shift}, 
it is natural to evaluate it by treating the long-range interactions
in the atom pair as perturbation of the pure isotropic harmonic
oscillator states. To first order, the van der Waals interaction $-1/x^6$ gives rise to a contribution proportional to $\beta_\omega^6$, whereas the anisotropic term $-c_3/x^3$ results in the dominant contribution to the energy shift. It is proportional to $\beta_\omega^3$ and negative for $\ell$=1, $m$=0 or $\ell\ge 3$, $|m|=0, 1$, when the adiabatic potential is attractive, and positive for $\ell$=$|m|$=1, when the potential is repulsive. The full  expression of the dominant term of $A$ is reported in Table~\ref{tab:trap}.
%
%%%%%%%%%%%%%%%%%%%%%%%%%%%%%%%%%%% TABLE VI %%%%%%%%%%%%%%%%%%%%%%%%%%%%%%%%%%%%%%%
% Energy shift of the trap levels
%%%%%%%%%%%%%%%%%%%%%%%%%%%%%%%%%%%%%%%%%%%%%%%%%%%%%%%%%%%%%%%%%%%%%%%%%%%%%%%%%%%%
\begin{table}[tb]
%\small
  \begin{tabular}{|c|c|c|c|}
    \hline
				%\multicolumn{2}{|c|}{trap level}  & \multicolumn{2}{c|}{numerical fit}& \multicolumn{2}{c|}{analytical }   \\
$N$ &$\mathcal E_0$&$A$ &$B$   \\                      \hline \hline
$0$	   & $5 \beta_\omega^2$    &$-\frac{4\beta_\omega^3 c_3}{3\sqrt\pi}$&$\frac{8\beta_\omega^5 }{\sqrt\pi}  $ \\                     \hline
$1$	   & $9 \beta_\omega^2$    &$-\frac{26\beta_\omega^3 c_3}{15\sqrt\pi}$&$\frac{20\beta_\omega^5 }{\sqrt\pi} $\\                  \hline
$2$	   & $13 \beta_\omega^2$    &$-\frac{433\beta_\omega^3 c_3}{210\sqrt\pi}$&$\frac{35\beta_\omega^5 }{\sqrt\pi}$  \\               \hline
  \end{tabular}
  \caption{\label{tab:trap} Parameters $A$ and $B$ of Eq.~\eqref{eq:osc-shift}. $\Delta {\mathcal E}_{N,\ell=1,m}$ is the energy shift from the unperturbed energy ${\mathcal E}_0$=$2 \beta_\omega^2 (2N+5/2)$ of a trapped $\ell$=1 state of a pair of particles submitted to a non-resonant light of reduced intensity $\mathcal I$ and $\mathcal M_0^m(x_{00})$ is the field-dressed generalized scattering volume of the pair when untrapped. The $m$- and ${\mathcal I}$-dependences are those of the $c_3$ coefficient of the adiabatic $\ell$=1, $m$ field dressed potential $c_3$=$4 {\mathcal I}/15$ ($-2 {\mathcal I}/15$) for $m$=0 ($|m|$=1). All data are in reduced units. }
\end{table}
%%%%%%%%%%%%%%%%%%%%%%%%%%%%%%%%%%%%%%%%%%%%%%%%%%%%%%%%%%%%%%%%%%%%%%%%%%%%%%%%%%%%%
%
A straightforward analytical evaluation of the parameter $B$ is obtained 
by representing the short-range interactions for each partial wave by a contact potential, with strength proportional to the energy-dependent scattering parameter 
${\mathcal S}_{\ell}(E)=(a_\ell(E))^{2\ell+1}$
for the corresponding $\ell$-wave collision~\cite{Derevianko05,Idziaszek06}.
In the single-channel approximation, the energy of the trapped bound
levels is related to the scattering parameter ${\mathcal S}_{\ell}(E)$
by an implicit transcendental $\ell$-dependent equation involving reduced units of the harmonic oscillator, cf. Eqs.~\eqref{eq:ru-osc-x} and~\eqref{eq:ru-osc-e}, 
\begin{equation}
\label{eq:new}
{\mathcal S}_{\ell}(E)/(a_\omega)^{2\ell+1}= %\underline{f_{\ell}}(E/\epsilon_\omega)=
\underline{f_{\ell}}(e_\omega)\,,
\end{equation}
where $\underline{f_{\ell}}(e_\omega)$ is expressed analytically in terms of $\Gamma$-functions and depends only on the reduced energy $e_\omega$~\cite{Busch98,Bolda02,Kanjilal04,Kanjilal07}. Equation~\eqref{eq:new}  implicitly connects the exact trap state energy of the particles, that interact via an energy-dependent short-range interaction, to the scattering parameter. For scattering in tight traps, it is essential to 
introduce energy-dependent scattering parameters since the Wigner threshold law may not apply at a given trap energy \cite{Bolda02}. 

Equation~\eqref{eq:new} has to be solved self-consistently for each eigenenergy. If we consider, for example, 
${\mathcal S}_{\ell=1}(E)$=0, which corresponds to vanishing
short-range interactions in the $p$-wave, Eq.~\eqref{eq:new} possesses
several roots $E_{N,\ell=1, m}$, each one associated with a state of
the unperturbed  isotropic 3D harmonic oscillator level $e_\omega$=$2N+\ell+3/2\,$ru($\omega$). A value of the parameter $B$, which accounts for the short-range interaction to first order in perturbation theory, 
is analytically obtained from the derivative of the function $\underline{f_{\ell}}(e_\omega)$ for a vanishing value of the scattering parameter. In van der Waals reduced units, one has 
\begin{eqnarray}
B&=&\frac{dE}{d ({\mathcal S}_{\ell=1}(E))}\,\Bigg|_{\,{\mathcal S}_{\ell=1}(E)=0}
\nonumber
 \\ &=&
2 \beta_\omega^{\;5} \frac{de_\omega}{d 
{\underline{f_{\ell=1}}}(e_\omega))}\,\Bigg|_{\,e_\omega=2N+5/2} \,. 
\label{eq:B}
\end{eqnarray}

The values obtained for $B$, which vary as $\beta_\omega^{\;5}$, are reported in Table~\ref{tab:trap}. For the lowest trap level, the resulting energy shift 
is identical to Eq.~\eqref{eq:AB-LK}. For $N$=2 and all other
parameters as in Fig.~\ref{fig:trap-0} (resp. Fig.~\ref{fig:trap-1}),
the shift $A$ becomes
 $-0.0002327$~ru (resp. $0.0001163$~ru), to be compared to -0.00026685~ru (resp. 0.000111~ru) as quoted in the figure captions. The multiplicative factor $B=6.17\times 10^{-6}$~ru, which is the same for $m$=0 and $|m|$=1, has to be compared to the value of $5.9\times 10^{-6}$~ru in the two figure captions.

\section{Dependence of the relative orientation at short range on the nodal parameter $x_{00}$}
\label{sec:eta}

%
% %%%%%%%%%%%%%%%%%%%%%%%%%%%%% fig 10 rapport des pentes %%%%%%%%%%%%%%%%%%%%%%%%
%
\begin{figure}[tb]
  \centering 
  \includegraphics[width=0.99\linewidth]{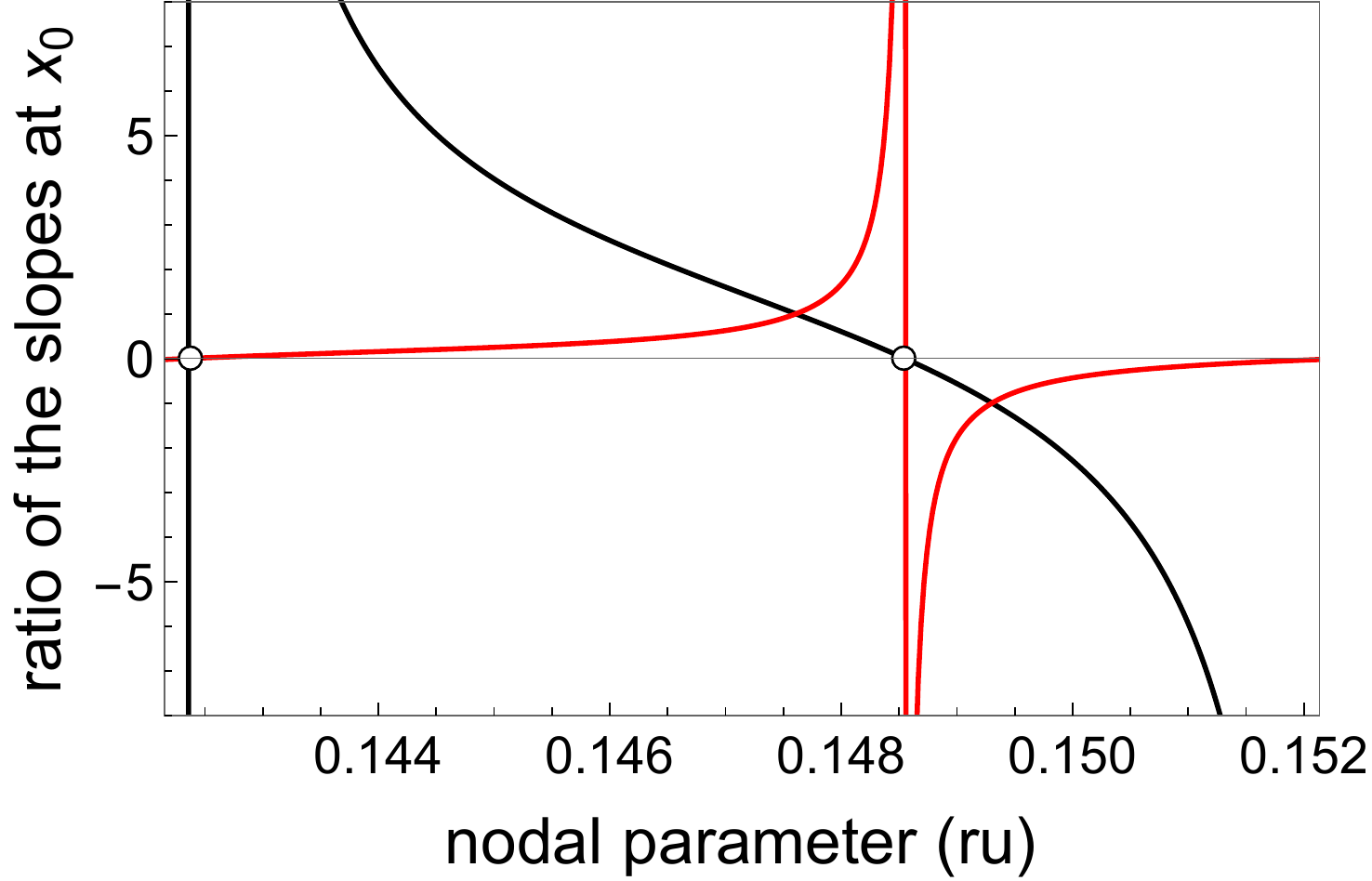}
  \caption{Nodal parameter dependence  of     $D_0(x_0)/D_1(x_0)$,
    i.e., the ratio of the slopes of the $m$=0 and the $|m|$=1 wave
    functions at    the position of their repulsive wall $x_0(\mathcal E=0, \ell=1, \mathcal I)$ (in black), together with the
    $x_{00}$-dependence of the inverse ratio (in red). The intensity is ${\mathcal I}$=6~ru. This ratio is independent of the asymptotic main orientation $\alpha$. The two small open circles indicate the positions of the divergences of the generalized scattering volume for $m$=0 (left) and $|m|$=1 (right). }
  \label{fig:ratio-of-slopes}
\end{figure}
%%%%%%%%%%%%%%%%%%%%%%%%%%%%%%%%%%%%%%%%%%%%%%%%%%%%%%%%%%%%%%%%%%%%%%%%%%%%%%%%
%

The $x_{00}$-dependence of the limit of $\eta (x)$ for $x\rightarrow x_{00}$, shown in Fig.~\ref{fig:orientation}
in Sec.~\ref{sec:orient},
can be also understood by calculating the slopes
$D_m(x_{0})=u'_m(x_{0})$ of the two solutions $u_{m=0,\pm 1}$
at the position of their energy-, intensity- and $\ell$-dependent
repulsive wall, $x_{0}(\mathcal E=0,\ell,\mathcal I)$ (defined in App. C.2 of Paper I~\cite{paperI} and Ref.~\cite{CrubellierNJP15a}).
This is shown in Fig.~\ref{fig:ratio-of-slopes}, where the $x_{00}$
dependence of the  ratio  $D_0(x_{0})/D_1(x_{0})$  
and that of its inverse are presented. 
These ratios are independent of  $\alpha$, the asymptotic main orientation.
A divergence of the ratio $D_0(x_{0})/D_1(x_{0})$ appears when the
generalized scattering volume diverges for $m$=0. This is due to the rapid variation of the amplitude of the oscillations of $u_0$
with $x_{00}$ in the inner region, associated with a divergence of
$D_0(x_{0})$, the slope of the function $u_0(x)$ at the position $x_0$
of the repulsive wall, 
 and is a signature of the presence of a bound state at threshold for $m$=0. The normalized wave function of this bound state has then a very large amplitude in the inner region, as is the case for a shape resonance.
This agrees with the results of the Levy-Keller model using the BC2 reference functions, cf. Paper I~\cite{paperI}. In this model one has $u(x)\propto x^2 - {\mathcal M_{BC2}(x)}/x$. For small $x$, the $1/x$ contribution prevails 
and, when $x_{00}$ varies, the short range amplitude of $u(x)$ and the
generalized scattering volume $\mathcal M^0_{BC2}$ diverge for the
same $x_{00}$ value. Analogously, and for similar reasons, a
divergence in $x_{00}$ of the ratio $D_1(x_{0})/D_0(x_{0})$ appears
when the generalized scattering volume $\mathcal M^0_{BC2}$ diverges
for $|m|$=1, associated with an $|m|=1$-bound state  at threshold.

%---------------------------------------------------------------------------------------
\bibliography{shaperes}
%---------------------------------------------------------------------------------------
\end{document}